%% file: paper-arxiv.tex
\documentclass{sig-alternate}
\usepackage{graphicx}
\usepackage{array}
\usepackage{xspace}
\usepackage{outlines}
\usepackage[hyphens]{url}
\usepackage[usenames,dvipsnames]{color}
\usepackage{balance}
\usepackage{footnote}
\usepackage{etoolbox}
\makeatletter
\AtBeginEnvironment{tabular}{%
  \@currsize}%
\makeatother
\usepackage{algorithm}
\usepackage{algpseudocode}
\usepackage[font={small,bf}]{caption}
\DeclareCaptionLabelFormat{andtable}{#1~#2  \&  \tablename~\thetable}
\makesavenoteenv{tabular} 
\usepackage{subcaption}

\input{macros}

\begin{document}

\numberofauthors{1}
\begingroup
\title{View-Driven Deduplication with Active Learning}
\author{
  \alignauthor
  Kristi Morton, Hannaneh Hajishirzi, Magdalena Balazinska, and Dan Grossman \\
\affaddr{Department of Computer Science and Engineering \\
University of Washington, Seattle, WA, USA} \\
   \email{{\small \{kmorton, hannaneh, magda, djg\}@cs.washington.edu}}
}
\endgroup
\maketitle
\begin{sloppypar}
\begin{abstract}
  Visual analytics systems such as Tableau are increasingly popular
  for interactive data exploration. These tools, however, do not
  currently assist users with detecting or resolving potential data
  quality problems including the well-known deduplication
  problem. Recent approaches for deduplication focus on cleaning
  entire datasets and commonly require hundreds to thousands of user
  labels.  In this paper, we address the problem of deduplication in
  the context of visual data analytics.  We present a new approach for
  record deduplication that strives to produce the cleanest view
  possible with a limited budget for data labeling.  The key idea
  behind our approach is to consider the impact that individual tuples
  have on a visualization and to monitor how the view changes during
  cleaning. With experiments on nine different visualizations for two real-world datasets, 
  we show that our approach produces significantly cleaner views for
  small labeling budgets than state-of-the-art alternatives and that
  it also stops the cleaning process after requesting fewer labels.

  
\end{abstract}



\input{intro}

\input{problem}
\input{background}
\input{approach}
\input{experiments}

\input{related}

\input{conclusion}
\input{acks}
\end{sloppypar}

\scriptsize
\bibliographystyle{abbrv}
\sloppy
\bibliography{vldb17}

\end{document}

%% file: macros.tex
\newcommand{\leaveout}[1]{}

\newtheorem{EXAMPLE}{View}
\newcommand{\xam}{\begin{EXAMPLE} \hspace{-.25em}}
\newcommand{\exam}{\vspace{0.1in} \end{EXAMPLE} }

\newtheorem{definition}{Definition}[section]   
 \newcommand{\defn}{\begin{definition} \hspace{-.25em}}
\newcommand{\edefn}{\vspace{0.01in} \end{definition} }

\newtheorem{EXAMPLE1}{Example}

\newenvironment{packed_item}{
\begin{itemize}
   \setlength{\itemsep}{1pt}
   \setlength{\parskip}{0pt}
   \setlength{\parsep}{0pt}
}
{\end{itemize}}

\newenvironment{packed_enum}{
\begin{enumerate}
   \setlength{\itemsep}{1pt}
   \setlength{\parskip}{0pt}
   \setlength{\parsep}{0pt}
}
{\end{enumerate}}

\newcommand{\me}{Many Eyes\xspace}  

\newcommand{\ie}{{\em i.e., }}
\newcommand{\eg}{{\em e.g., }}

\newcommand{\eat}[1]{}



%% file: intro.tex
\section{introduction}
\label{s:intro} 

Visual analytic systems such as Tableau~\cite{vizql} are becoming
increasingly popular for data exploration and analysis. These tools
enable users to interactively query data through a drag-and-drop
interface, and the results are rendered on-the-fly as visualizations.
These visualizations are represented internally as database
views. Users can create a collection of sophisticated views that combine multiple
heterogeneous data sets (\eg Excel spreadsheets, relational databases,
data cubes, delimited text files, etc.) along a common dimension or
set of dimensions.

Today's visual analytics systems assume that the data sets being
consumed are clean and consistent with respect to each other (\eg all
entities in canonical form). However, data (especially on the Web) is
often subject to data quality problems. Deduplication is one kind of
dirty data problem. This problem manifests when there are different
representations of the same real world entity or object in the data
sources being integrated. For example, the same restaurant
may appear under two different phone numbers.  The same product may use
different abbreviations in its name or may include a different
description.

\begin{figure}
\begin{center}
\includegraphics[width=1\linewidth]{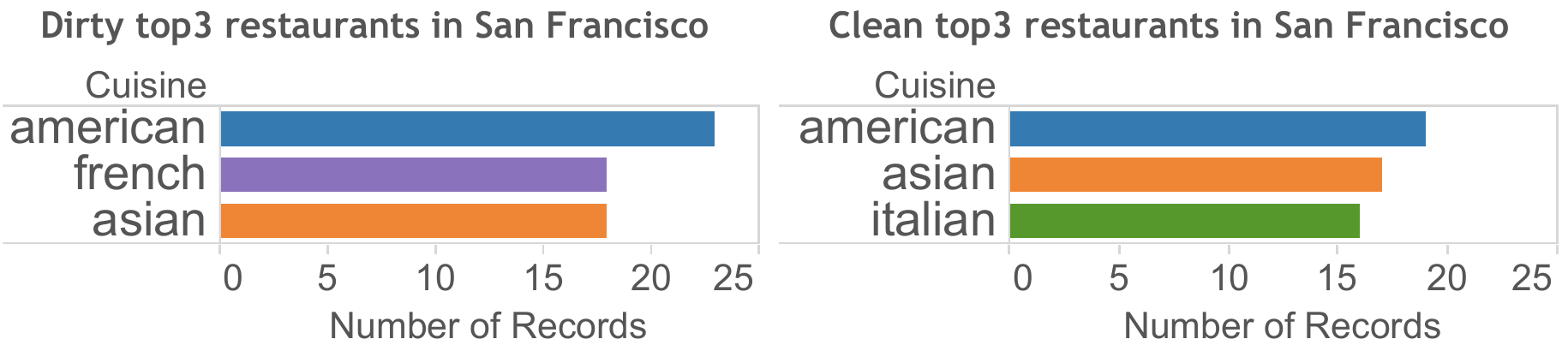}
\vspace{-.7cm}
\caption{Views, dirty (left) and cleaned after 25 labels (right), over Fodor $\cup$ Zagat restaurant datasets.} 
\vspace{-.6cm}
\label{fig:top-restaurants-view}
\end{center}
\end{figure}

Duplicate records may affect a
visualization. Figure~\ref{fig:top-restaurants-view} shows an example,
which we use as running example throughout the paper. The figure shows
the top three types of cuisines by quantity of restaurants in San
Francisco. This view is computed over a restaurant dataset commonly
used for the evaluation of entity resolution tools~\cite{riddle}. This
dataset was created from the union of the Fodor and Zagat restaurant
ratings datasets.  Because this is a benchmark dataset, duplicate
tuples are labeled as such. We can thus compute the view both over the
dirty and clean versions of the data. The figure shows the two
resulting visualizations.  Duplicate records clearly impact the
visualization: French restaurant records are incorrectly
overrepresented in the view on the left and should therefore be
removed (or otherwise merged). The problem, however, is that data cleaning is a disruptive
process. It interrupts the user during his primary data exploration
task. Our goal is to clean a user's visualization (or set of
visualizations) with minimal interruption, \ie minimal number of
requests to the user for assistance.

In this paper, we focus on the entity resolution problem. Given a view
computed over a dataset that contains duplicate entities, our goal is
to \textit{clean} the view by identifying and removing duplicate
records in the underlying dataset. The commonly used approach to
deduplicating a set of records, $R$, comprises the following steps. The
process requires as input a similarity function that takes a pair of
tuples $(t_1,t_2)$ from $R$ and produces a similarity
score~\cite{elmagarmid2007duplicate}. This similarity function,
together with a similarity threshold, is applied to all pairs of
records in $R$ to determine which ones match~\cite{dey1998entity}.
Alternatively, a system may rely solely on users to indicate which
tuples match~\cite{kang2007geoddupe,bilgic2006d}. Multiple tuples can
correspond to one entity and such clusters further need to be
identified~\cite{altwaijry:2013query,bhattacharya2007query,wang2013leveraging,vesdapunt2014crowdsourcing,wang2015crowd}. Once
matching tuples are identified, they must be
merged~\cite{guha2004merging,Bleiholder:2009}. During this step, any
data conflicts among the multiple representations must be
resolved. This step is called data fusion~\cite{Bleiholder:2009} and
is not covered in this paper. To make the previous steps more
compute-efficient by reducing the number of record comparisons,
blocking techniques are used~\cite{bilenko2006}. Blocking is an
inexpensive heuristic filtering step that either partitions the tuples
that get compared or removes pairs with low similarity scores.

Since manually devising an accurate similarity function requires an
expert, state-of-the-art techniques for deduplication use active
learning instead~\cite{mozafari2015scaling, gokhale:2014corleone},
where one or more users label training examples (\ie pairs of tuples)
as either duplicates or not, which enables the system to learn a
classifier that categorizes the remaining pairs of tuples.  Active
learning iteratively asks users for additional, carefully selected
labels and re-trains the classifier until the classifier stops
improving.

Active-learning-based deduplication is a promising approach for
cleaning data visualizations.  As a simple example, consider a typical
data enthusiast, a food journalist, who wants to publish some
visualizations that tell a story about restaurants in San Francisco by
the end of the workday. After downloading a US restaurant ratings
dataset from the Web that has duplicate entities and before visually
exploring it using Tableau (or some other system), the journalist may
choose to clean the data. The active learning
method~\cite{Bellare:2012,arasu:2010active,beygelzimer2009importance} would
select pairs of records and would ask the user to label
them as either duplicates or not. It would then use the labels
to build a classifier. Active learning repeats the process until
the classifier stops improving. The classifier then labels
all remaining pairs. After the classification completes, matching
records can be merged to yield the clean dataset.



Existing active learning methods produce high-quality classifiers, but
at great cost to the user. The user may have to provide hundreds to
thousands of labels during the data cleaning process (see Table 2
in~\cite{gokhale:2014corleone} and Figure~\ref{all-results-products}
in Section~\ref{sec:eval}), which is significant for a data enthusiast
who most likely just wants to create one or a few
visualizations~\cite{morton2014public}.  Several systems use the crowd
to perform the cleaning~\cite{gokhale:2014corleone}, but that approach
takes days to complete, which is also inconsistent with our data
enthusiast scenario.


In this paper, we develop an approach that addresses the above
problem. We present a new active-learning-based method to
classify tuples in a view as either duplicates or not. In contrast to prior work
mentioned above, our approach focuses on producing the cleanest
view or set of related views (and not necessarily the cleanest dataset) with a
small budget of labels from the user (no crowd). We assess
\textit{view cleanliness} by computing the distance of the current
view to the view over the dataset with all duplicates removed.


Our method,\textit{View Impact Cleaning}, performs
deduplication in a manner that focuses on a user's current visualization
(or a set of related visualizations, as in a dashboard). View Impact Cleaning
yields a significantly cleaner view than active learning alone
when given a small labeling budget. It only asks the user
to label data that is currently being visualized and 
stops the cleaning process when it determines that additional
labels will not change the visualization further even if they could yield
a better overall classifier.

By developing the View Impact Cleaning method, our contributions include:

\begin{packed_enum}
\item A new notion of \textit{view sensitivity} to duplicate
  tuples. View sensitivity captures the extent to which a view is
  affected by duplicate tuples. We also define a new notion of
  \textit{view impact score} of individual tuples on a
  visualization. The view impact score measures the extent to which a
  view will change if a given tuple is found to be a duplicate and is
  removed  (Section~\ref{sec:sensitivity}).

\item An active-learning method that builds an
  initial classifier and then iteratively improves that
  classifier.  One
  novelty of our approach is in the selection of the training
  examples: it considers both the view impact scores of
  individual tuples and the potential of a training example to improve
  the classifier quality (Section~\ref{sec:cleaning}).

\item A new stopping condition for view cleaning that
considers the view's evolution during the cleaning process. An important
implication of our approach is that it stops cleaning a view
both in the case where a sufficient number of tuples have
been removed and in the case where a view is not sensitive to
duplicate tuples and cleaning has little effect on the view (Section~\ref{sec:stop}).

\end{packed_enum}

We evaluate our approach on nine views specified on two
real-world entity resolution data sets. We use the restaurants dataset from the well-studied RIDDLE~\cite{riddle} repository and the Google-Amazon products dataset from~\cite{amazon-google-products,kopcke:2010,kopcke2010learning}. We find that, when given a small cleaning budget (\ie
the user is willing to label a small number of record pairs as
duplicates or not), our approach yields significantly cleaner views
than existing active learning methods, which do not consider the
users's view (or dashboard of views). It also effectively stops cleaning earlier than active
learning alone while delivering views much closer to those computed
over the clean data. Finally, we evaluate and discuss the problem of
cleaning a dashboard comprising multiple visualizations. Our results
show that cleaning one view with our approach effectively helps to
clean other views even though cleaning is view-driven. As such, our
approach helps to make data cleaning a pay-as-you-go task.

%% file: problem.tex

\section{Problem Statement}
\label{s:problem}

Consider a relation $R$ that contains duplicate tuples. Two
tuples $t_1$ and $t_2$ $\in$ $R$ are
duplicate if they refer to the same real-world entity (\eg same
restaurant).  They need not be identical and, most often, are not
identical. For
example, the same restaurant may appear twice but with different phone
numbers.  The relation may be the result of the integration of two or
more datasets or may contain duplicate tuples for other reasons. We
assume $R$ to be given and we do not require knowledge of where
individual tuples in $R$ come from. In our running example from
Figure~\ref{fig:top-restaurants-view}, $R$ is the unioned restaurant
dataset.

The user builds a view, $V(R)$, and a visualization that displays it,
such as the one shown in Figure~\ref{fig:top-restaurants-view}. In our
approach, we do not
consider the details of the visualization. Instead, we focus on
the relation $V(R)$ and consider that any change to $V(R)$ affects the
visualization.  Our approach supports views that correspond
to select-project queries with optional aggregation, grouping,
sorting, and top-K restrictions.

We define $R_{clean}$ as the relation $R$ with all duplicate tuples removed. For
convenience, we refer to the original view, $V(R)$, as $V_{dirty}$, to
$V(R_{clean})$ as $V_{clean}$, and to the same view $V$ computed on a
partially cleaned relation as $V_{curr}$. We define the
\textit{quality}, or \textit{cleanliness}, of a view as follows:

\defn{\texttt{Quality(V)} or \texttt{Cleanliness(V)} of a view $V$ is \texttt{1 - Distance(V,
    $V_{clean}$)}, for some distance function, \texttt{Distance} $\in [0,1]$. }\edefn

The quality of a view thus depends on the distance to the view
computed on clean data.  In our implementation, we use the well-known
Earth Mover's Distance~\cite{rubner:2000earth} to compute distances between views as we
describe in Section~\ref{sec:sensitivity}.

\textbf{Objective:} The goal of our approach is to clean a view by
reducing $\mathrm{Distance}(V, V_{clean})$.

We target scenarios where a single user explores a
dataset $R$ by defining one or more views on top of $R$. We assume
that the user is a data enthusiast who can label pairs of records in
his dataset as either duplicate or not but cannot otherwise tune or
help the data cleaning process.

In this paper, we do not address the problem of how best to merge
duplicate tuples~\cite{Bleiholder:2009}. Any of the existing
techniques~\cite{getoor:2012entity} could be used. In our experiments,
we drop one of the duplicate tuples.  We also do not handle
labeling errors~\cite{mozafari2015scaling}. We assume correct
labels. These additional techniques are complementary to the approach
developed in this paper. Additionally, our approach relies on the data
enthusiast to provide labels directly. We do not use the crowd. We do
not require \textit{any} expertise from the user beyond the ability to
identify whether two records are duplicates.

%% file: background.tex
\section{Background}
\label{sec:background}



Deduplication has been a long-standing, challenging
problem~\cite{halevy2012principles, getoor:2012entity}.  The most
closely related work applicable to our context relies on a non-expert
user or users to label tuple pairs as either duplicate or not and then
uses machine learning to build a classifier to identify duplicate
records in a
relation~\cite{arasu:2010active,Bellare:2012,gokhale:2014corleone,
  kopcke2008training,mozafari2015scaling,vesdapunt2014crowdsourcing,whang2013question}. We
build on this foundation, which we briefly review here:


\textbf{Learning a Classifier:} Given the relation $R$ to clean, one
builds a cartesian product $S = R \times R$ (\eg $S$ is the set of
all restaurant pairs). For each tuple in $S$, one computes a feature
vector that captures distance information between the individual
attributes of the two $R$ tuples that form the $S$ tuple. For example,
one feature could be the edit distance between restaurant names and a second the
jaccard similarity between their addresses. Commonly used distance
functions include edit distance, jaccard, jaccard containment, and
cosine distance
(see~\cite{cohen2003comparison,elmagarmid2007duplicate} for detailed
descriptions) for string attributes and Euclidean distance for
numerical attributes. Other functions are
possible~\cite{cohen2003comparison}.



The basic learning algorithm selects a random sample of pairs
from $S$, asks the user to label them as either duplicates or not, and then
learns a classifier using that training data. 

Because $|S|$ can be large and because the number of positive examples
is typically small compared with the number of negative examples, a
blocking function serves to reduce $|S|$ before the selection of the
training examples. The blocking function takes the form of a selection
predicate on $S$, where the predicate retains only pairs with distance
between specific attributes below a pre-defined threshold. For
example, a blocking function can retain only pairs of restaurants whose
names have an edit distance below some threshold.

\textbf{Active Learning:} Active learning improves on the above
approach by iteratively training classifiers on increasingly large and
carefully selected training examples. As above, the initial step is to
learn a classifier on a random sample of training examples. Active
learning then selects additional training examples with the purpose of
improving the classifier. Several methods exist to select the
additional examples that are most informative. Current methods to
assess the \emph{informativeness} of training examples measure the
disagreement among a set of component classifiers using
uncertainty~\cite{mozafari2015scaling} or
entropy~\cite{gokhale:2014corleone}. Each component classifier is
trained on a random sample (with replacement) of the original training
dataset. The intuition is that the more the component classifiers
disagree on the predicted label for an example pair, the more likely
that the classifier will learn something new from this example and
thus produce an improved classifier. Whatever the method, active
learning then retrains a new classifier and repeats the
process. Learning stops when the classifiers stop improving across
iterations.

\textbf{Our Approach:} In this paper, we develop a new
active-learning-based deduplication approach. In contrast to the prior
work above, our approach cleans a user's view or set of views more
effectively and with fewer user labels. We evaluate cleaning effectiveness
by computing the distance between a user's view and the same view
computed over perfectly clean data. We present the details of our
approach in the next section and quantitatively compare it against
the above prior work in Section~\ref{sec:eval}.






%% file: approach.tex

\section{Approach}
\label{s:approach}

The goal of our approach is to deduplicate $R$ in a way
that minimizes the distance between $V_{curr}$ and $V_{clean}$
while keeping the number of tuple-pairs that the user needs to label
low. We employ an active-learning-based approach with the same
fundamental setting as presented in Section~\ref{sec:background}.



To clean $V(R)$, our approach is to build a classifier that takes as
input all pairs of tuples $(t_1,t_2)$ with
$t_1 \neq t_2 \wedge t_1 \in R \wedge t_2 \in R$ and classifies each
one as either a duplicate pair or a non-duplicate pair. Once
duplicates are identified, any existing method can serve to merge them
as indicated in Section~\ref{s:problem}. Our goal is to produce the
cleanest possible view (\ie smallest Distance$(V_{curr},V_{clean})$)
for a given label budget $l$. Additionally, we require that $l$ be
small in the order of tens or low hundreds of labels.




In this section, we first describe our model to reason about the
sensitivity of a view to duplicate tuples and the impact of individual
tuples on the view (Section~\ref{sec:sensitivity}). We then
present our active-learning approach to view cleaning, which is based
on this model (Sections~\ref{sec:cleaning} and~\ref{sec:stop}).




\subsection{View Impact Score and View Sensitivity}
\label{sec:sensitivity}







Our data cleaning approach introduces and leverages two important
concepts related to the way tuples affect a view. We call them
the \textit{view impact score} and \textit{view sensitivity}.

The \textit{view impact score} of a tuple measures how much the tuple
affects a view $V(R)$. We define it as follows:

\defn{The \texttt{view impact score} of a tuple $t \in R$
on a view $V(R)$ denoted as \texttt{Impact(V,t)} is \texttt{Distance(V(R),V(R-t))}
\label{def:tuple-impact}
}\edefn

The view impact score of a tuple measures the distance between the
view computed over the base relation $R$ with the tuple included and
with the tuple removed. View impact scores drive the cleaning process.
By identifying tuples with high view impact scores, our approach
effectively focuses cleaning actions on the subset of $R$ that
matters the most to the user's view(s).  For example, consider the top-k
view of cuisine types by quantity of restaurants in
Figure~\ref{fig:top-restaurants-view}: a tuple whose \texttt{cuisine}
attribute is `American', `Asian', `French', or `Italian' would have a
higher impact score because these cuisines appear in the view, than
one with a rare type such as `Indonesian', because the latter will
never appear in the view, even once the base data is completely clean.

The second concept, \textit{view sensitivity}, serves to inform
when to stop the cleaning process. The view sensitivity
measures how much a view is affected by duplicate tuples:

\defn{The \texttt{sensitivity} of a view $V_{curr}$ to duplicate
  tuples is \texttt{Distance$(V_{curr},V_{clean}$)} 
\label{def:sensitivity}
}\edefn

The sensitivity of a view is the distance between the current view and
the same view computed over the cleaned relation $R_{clean}$.  A view
is no longer sensitive to duplicate tuples for one of two reasons:
Either relation $R$ has been sufficiently cleaned or the view is
generally not affected by duplicate tuples. In both cases, any further
cleaning will not change the view in a significant way. For example, a
view that displays median values is not easily affected by duplicate
results: cleaning is unlikely to affect the median value
significantly. As another example, once the view from
Figure~\ref{fig:top-restaurants-view} correctly lists the top three
types of cuisine any further data cleaning will at most cause small
changes in the detailed counts for each cuisine type. The view
sensitivity at that point will be small.


During the view cleaning process, the system does not have access to
$R_{clean}$ and thus $V_{clean}$. Instead, our approach operates on
distances between consecutive views obtained during the iterative view
cleaning process, \texttt{Distance($V_{curr}$,$V_{curr+1}$}), to
estimate sensitivity and determine when to stop cleaning. 

\begin{figure}[t!]
\begin{center}
\tabcolsep=0.11cm
\includegraphics[width=\linewidth]{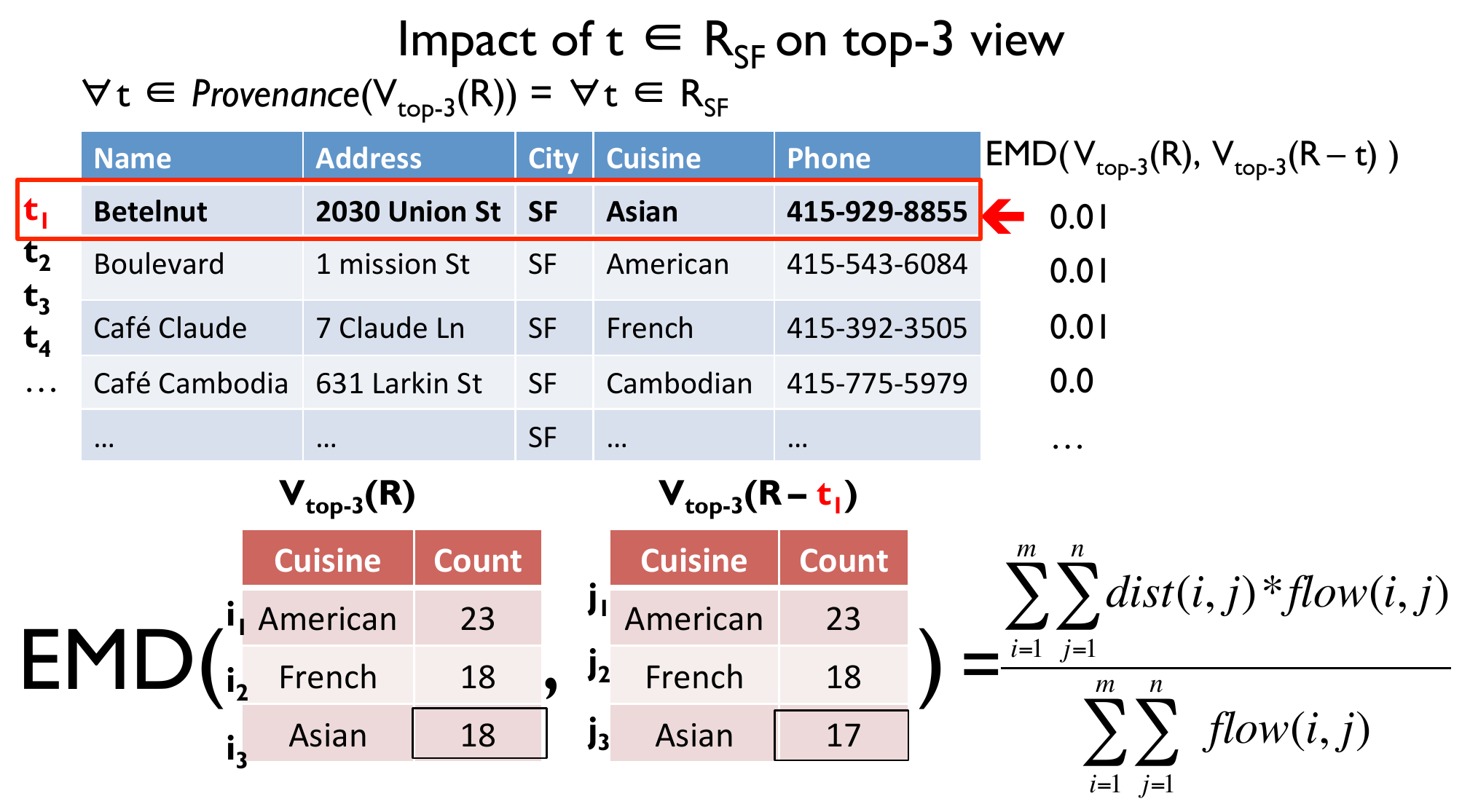}
\vspace{-0.5cm}
\caption{Example view impact score and view distance
  computation. (top) The view impact score for each tuple $t$ is the
  distance (EMD)
between the view over relation $R$ and over $R$ with tuple $t$ removed. 
(bottom) Illustration of one distance (EMD) computation between two views.}
\vspace{-0.5cm}
\label{fig:emd-views}
\end{center}
\end{figure}

\begin{table*}[t!]
\begin{small}
\begin{scriptsize} 
\begin{center}
\begin{tabular}{|l|l|c|c|c|} \hline
${i \in V(R_{SF})}$ & ${j \in V(R_{SF} - t_1)}$ & Attribute dist($i$,$j$) = & Tuple dist($i$,$j$) = & flow($i$,$j$)\\ 
{[$i_{Cuisine}$, $i_{Count}$]} & {[$j_{Cuisine}$, $j_{Count}$]}  & [1-StrEq($i_{Cuisine}$,$j_{Cuisine}$),normEuclid($i_{Count}$,$j_{Count}$)] &Euclid(Attribute dist($i$,$j$))& \\ \hline
\bf{[``American", 23]} & \bf{[``American", 23]} & \bf{[ 0.0,  \ \ \ \ \ \ \ \ \ \ \ \ \ \ 0.0 ]} & \bf{0.0} & \bf{1/3} \\  
{[``American", 23]} &  {[``French", 18]} &  {[ 1.0, \ \ \ \ \ \ \ \ \ \ \ \ \ \ 0.217 ]} & 1.023 & 0.0 \\
{[``American", 23]}&  {[``Asian", 17]}& {[ 1.0, \ \ \ \ \ \ \ \ \ \ \ \ \ \ 0.261 ]}  & 1.033 & 0.0 \\
{[``French", 18]} &        {[``American", 23]} & {[ 1.0, \ \ \ \ \ \ \ \ \ \ \ \ \ \ 0.217 ]} & 1.023 & 0.0 \\
\bf{[``French", 18]} &        \bf{[``French", 18]} & \bf{[ 0.0,   \ \ \ \ \ \ \ \ \ \ \ \ \ \ 0.0 ]} & \bf{0.0} & \bf{1/3} \\
{[``French", 18]} &    {[``Asian", 17]} & {[ 1.0, \ \ \ \ \ \ \ \ \ \ \ \ \ \ 0.043 ]} & 1.001 & 0.0 \\                   
{[``Asian", 18]} &       {[``American", 23]} & {[ 1.0, \ \ \ \ \ \ \ \ \ \ \ \ \ \ 0.217 ]} & 1.023 & 0.0 \\
{[``Asian", 18]} &      {[``French", 18]} & {[ 1.0,  \ \ \ \ \ \ \ \ \ \ \ \ \ \ \ \ \ 0.0 ]}  & 1.0 & 0.0 \\
\bf{[``Asian", 18]} &      \bf {[``Asian", 17]} & \bf{[ 0.0, \ \ \ \ \ \ \ \ \ \ \ 0.043 ]} & \bf{0.043} & \bf{1/3} \\ \hline\end{tabular}
\end{center}
\end{scriptsize}
\end{small}
\vspace{-.5cm}
\caption{EMD computation for
  ${V(R_{SF})}$ and ${V(R_{SF}-t_1)}$, from
  Figure~\ref{fig:emd-views}. $\forall$ pairs ($i$,$j$) $\in$
  ${V(R_{SF})}\times {V(R_{SF}-t_1)}$, dist($i$,$j$) is computed by
  applying the Euclidean distance to the set of attribute distances:
\eg the tuple distance for the
  third row in this table is: $\sqrt{(1.0)^{2} +
    (0.261)^{2}} \approx 1.033$. Each attribute has \emph{one}
  type-based distance function applied to it, \eg since \emph{Count}
  is a \texttt{num} type, the Euclidean distance is used (and
  normalized by the max value in the table so that the result is in
  the [0,1] range). For example, the $\texttt{normEuclid}(23,17)
  =\frac{\sqrt{(23-17)^{2}}}{23} \approx 0.261$. With the tuple
  dist($i$,$j$) and per-view tuple weights = 1/$|V|$ = 1/3 as input,
  we call the EMD library in~\cite{emdlib} to solve for the
  flow($i$,$j$) that minimizes movement of earth between the two views, or dist($i$,$j$)*flow($i$,$j$).} 
\vspace{-.3cm}
\label{feature-table}
\end{table*}

A key component of the above two concepts is the notion of
\textit{distance between two views}. Our approach requires a distance
function that captures differences at the level of tuples and
individual cells. One function that captures this requirement is the
Earth Mover's Distance (EMD)~\cite{rubner:2000earth}, which we
successfully applied in prior work~\cite{kwon2007identifying}. EMD is
a method to evaluate dissimilarity between two multi-dimensional
distributions.  Intuitively, given two distributions, one can be seen
as a mass of earth spread in space, the other as a collection of holes
in that same space. Then, the EMD measures the least amount of work
needed to fill the holes with earth. Here, a unit of work corresponds
to transporting a unit of earth by a unit of ground distance.



To compute the EMD between two views $V_1$ and $V_2$, we thus need a
distance function for individual tuples in these views and a weight
for each tuple. For the weight, we assign each tuple in a view $V$ the
same weight equal to $\frac{1}{|\mathrm{V}|}$.  For the tuple
distance, we consider each tuple t with \emph{n} attributes as an
\emph{n}-dimensional vector and use Euclidean distance to compute the
distance between two tuples $i \in V_1$ and $j \in V_2$. For
individual attributes in a tuple, we use Euclidean distance to compute
the distance between numeric attributes (normalized to the $[0,1]$
range) and string equality as the distance between categorical
attributes (\ie 0 if attributes are the same and 1 if they are
different). We explain the distance computation through an example:
Consider the following two views (in Figure~\ref{fig:emd-views}): (1)
$V_{top-3}(R)$, a top-3 view of cuisines in San Francisco over a
restaurants dataset, $R$, and (2) the same top-3 view but over a
relation $R'$ where one tuple $t$ has been removed.  The SQL statement
for this view appears in Table~\ref{all-view-sensitivity}.  The
distances are then calculated between all combinations of tuples
$(i,j)$ where $i \in V_{top-3}(R)$ and $j \in V_{top-3}(R - t)$, as
shown in Table~\ref{feature-table}. Each tuple is given the same
weight of $1/|V_{top-3}|$ and we call the library from~\cite{emdlib}
to solve the linear program that computes the minimum flow to move the
earth between the views using the pre-computed distances.  The
solution to the linear programming problem is shown bolded in
Table~\ref{feature-table}.  The EMD returned is
$\frac{0 * 1/3 + 0 * 1/3 + 0.043 * 1/3 } {1/3 + 1/3 + 1/3}$, which is
$0.01$. This value also corresponds to the view impact score of tuple
$t$ as per the definition above.

\subsection{View Cleaning}
\label{sec:cleaning}

In this section, we present our active-learning-based algorithm for
cleaning views by taking into account the view impact scores of
individual tuples and the view sensitivity to duplicates. The user
triggers the cleaning process, but stopping is automatic.




\subsubsection{Initial Classifier}  
\label{sec:cleaning:initial}

 
The first step in the active learning process is to select a set, $L_0$, of
training examples, ask the user to label them, and train an initial
classifier using those labels.  A training example is a pair of
tuples, $(t_1, t_2)$ with $t_1 \in R \wedge t_2 \in R \wedge t_1 \neq
t_2$. When a pair has duplicate tuples, it is a positive example.
Otherwise, it is a negative example. 

Recent prior work~\cite{mozafari2015scaling} randomly selects a 3\%
sample of such pairs to train the initial classifier. A known
challenge with record deduplication, however, is that the number of
positive examples is extremely small even when a dataset contains many
duplicate tuples. For example, if each duplicate tuple in a relation
of size $|R|$ participates in one positive example it also
participates in $|R|-2$ negative examples.  As a result, a small
random sample of training examples can easily fail to include any
positive examples, leading to a poor initial classifier, especially
when $|R|$ is large. A common approach to alleviate this problem is to
use blocking, where all tuple-pairs with low similarity scores for one
or more features are discarded before the data cleaning process even
begins. For example, pairs of restaurants with names that are not at
all similar should be discarded.  We further use a second blocking
method: We focus only on tuples that participate in the view. Instead
of cleaning |R|, we clean only those tuples in |R| that pass the
selection condition in the query. We denote these tuples with
\texttt{Provenance}$(V(R))$ since they correspond to the
why-provenance~\cite{buneman2001and} of $V(R)$ if we ignore any top-k
clauses in the query. Table~\ref{dataset-details} shows the fraction of duplicates for two
datasets that we describe further in Section~\ref{sec:eval}. The table
shows the result for both the dataset as a whole and for the subset of
the data in the views that we use in the evaluation.  Even after
applying both types of blocking (blocking on the view and the
features), the fraction of positive examples is only 2.3\% and 9.4\%
for the two views (we describe the exact blocking function in
Section~\ref{sec:eval}).



The second challenge with learning a classifier for record
deduplication is that the features themselves used to train the
classifier may be poor. In our application domain, in particular, the
user's goal is to create and analyze a given set of
visualizations. The user is not seeking to clean the data. As a
result, the system cannot rely on the user to determine a good set of
features. Instead, the feature selection process must be 
automated, which complicates the identification of a good set of
features. 




The above two challenges make it difficult to build high quality
classifiers as we show in the evaluation, and lead us to develop a
different strategy for training an initial classifier.  Our key idea
is to get the user to label tuple-pairs where at least one tuple has a
high view impact score. The intuition is that these pairs will not
necessarily be worse training examples than random pairs. At the same
time, correct labels for these pairs have the highest potential to
improve the quality of the view.  For example, in
Figure~\ref{fig:top-restaurants-view}, tuples that correspond to
American, French, Asian, and Italian restaurants will have higher view
impact scores than others and pairs containing such tuples should be
weighted more heavily when selecting examples to label.

The approach to learning the initial classifier has three main steps:
view impact score computation (Algorithm~\ref{algo:view-impact}),
training-example selection, and training of the initial classifier
(Algorithm~\ref{EMDalgorithm} lines 1 through 14). The view impact
computation proceeds as follows:

\begin{packed_enum}

\item For each tuple $t \in$ \texttt{Provenance}$(V(R))$, we compute
its view impact score, \texttt{score}, as per
Definition~\ref{def:tuple-impact}.  For example, in the view in
Figure~\ref{fig:top-restaurants-view}, we only compute the view impact
for restaurants in San Francisco. Other tuples necessarily have a view
impact score of zero. We store the results in a relation called
\texttt{TupleScores}.


\item For each tuple $t \in $ \texttt{TupleScores}, we generate
  $|$\texttt{Provenance}$(V(R))|-1$ potential training examples of the form $((t,
  u), \mathrm{score})$, where $u \in $ \texttt{Provenance}$(V(R)) -
  \{t\}$ and $\mathrm{score}$ is the view impact score for $t$.  We store the
  results in a relation called \texttt{PairScores}.
\end{packed_enum}

\noindent Selecting the initial training examples proceeds as follows:

\begin{packed_enum}
\item First, we apply a blocking function that removes obvious
  non-matches from the previously computed \texttt{PairScores}. The
  blocking function drops all pairs with at least one
  attribute that has a similarity below a pre-defined threshold (or
  distance above a pre-defined threshold). Section~\ref{sec:eval}
  describes the blocking function that we use in the experiments.
  This step corresponds to function \texttt{Block} on line 6 of
  Algorithm~\ref{EMDalgorithm}.

\item Second, we select $|L_0|$ examples from the filtered \texttt{PairScores} by
  using weighted random sampling with weights
  equal to the view impact scores.  To train a new classifier, we need a set of examples that are not only informative but also diverse~\cite{gokhale:2014corleone}, and hence as prior work~\cite{mozafari2015scaling, gokhale:2014corleone}, we use weighted random sampling rather than selecting the top-k pairs with highest view impact. This approach heavily weighs the
  pairs with high scores while randomly breaking ties between pairs
  that have the same score, such as pairs generated from the same
  initial tuple. This step corresponds to function \texttt{Select$_{bias}$} on line 7 of
  Algorithm~\ref{EMDalgorithm}. 

\end{packed_enum}

Finally, the user labels the selected pairs and these labels serve as
training examples for the initial classifier (lines 8 through 10 in
Algorithm~\ref{EMDalgorithm}).  The duplicate tuples identified
explicitly by the user or implicitly by the classifier are then removed from
the input data and the view is recomputed as shown in lines 11 and 12 in
Algorithm~\ref{EMDalgorithm}.

\begin{algorithm}[t]
\small
\caption{ViewImpactScores ($V(R)$)} 
\label{algo:view-impact}
\begin{algorithmic}[1]
\State {\small {\bf Input:} Base relation \textbf{$R$} and view \textbf{$V(R)$}}
\State {\small {\bf Output:} {\bf PairScores}, map of pairs to view impact scores}
\State {\small Let TupleScores = PairScores = $\emptyset$}
\For{\small {each tuple t $\in$ Provenance(V(R)) }}
\State {\small score = Distance($V(R)$, $V(R-t)$)}
\State  {\small TupleScores = TupleScores $\cup$ (t, score)}
\EndFor
\For {each pair (t, score) $\in$ TupleScores }
\For {each u $\in$ Provenance($V(R)$) - \{t\}}
\State PairScores = PairScores $\cup$ ((t,u), score)
\EndFor
\EndFor
\State {\small Return {\bf PairScores}}
\end{algorithmic}
\end{algorithm}


\begin{algorithm}[t!]
\small
\caption{{\small View Impact Cleaning(\emph{l},b,$b_{L0}$,$V(R)$)}}
\label{EMDalgorithm}
\begin{algorithmic}[1]
\State {\small {\bf Input: l} is the total labeling budget,}
\State {\small {\bf Input: b} is the batch size, $b_{L0}$ is $L_0$ size}
\State {\small {\bf $V_{dirty} = V(R)$} is the view to clean}
\State {\small {\bf Output:} $V_{curr}$ the current cleaned view}
\State {\small PS = ViewImpactScores($V_{dirty}$)} {\scriptsize // PS is PairScores}
\State {\small PS = Block(PS)} {\scriptsize // filter candidate pairs with blocking function}
\State {\small $L_0$ = $Select_{bias}$($b_{L0}$,PS)}  {\scriptsize // select pairs for user to label}
\State {\small $L_0$ = Label$(L_0)$}  {\scriptsize // label the $L_0$ pairs}
\State {\small PS = PS - $L_0$} {\scriptsize // remove labeled pairs from the score map}
\State {\small VL = $\theta_{L_0}$(PS)} {\scriptsize //train $\theta$ on $L_0$ \& label remaining pairs}
\State {\small dups = Matches($L_0$ $\cup$ VL)} {\scriptsize //get dups from $L_0$ \& VL}
\State {\small $V_{curr} = V(R -\{ \mathrm{dups} \} )$ } 
\State {\small view\_change = Distance($V_{curr},V_{dirty}$)} 
\State {\small \emph{l} = \emph{l} - $\mid$ $L_0$ $\mid$}    {\scriptsize   // remove user labeled pairs from budget}
\While{\scriptsize \emph{l} $>$ 0 \&  NOT Converged(view\_change)}
\State {\small$T$ = $Select_{top}$(b,PS)} {\scriptsize //top scoring pairs, applies tie breaker} 
\State {\small $T$ = Label$(T)$}  {\scriptsize // label the $T$ pairs}
\State {\small $T_{acc}$ = $T_{acc} \cup T$}  {\scriptsize // accumulate the $T$ pairs}

\State {\small PS = PS - T}  {\scriptsize // remove user-labeled pairs from PS}
\State {\small VL=$\theta_{L_0 \cup T_{acc}}$(PS)} {\scriptsize // train $\theta$ on ${L_0 \cup T_{acc}}$ \& label PS}
\State {\small dups = Matches($L_0$ $\cup$ VL)} {\scriptsize // get dups from $L_0$ \& VL}
\State {\small $V_{prev} = V_{curr}$}
\State {\small $V_{curr} = V(R -\{ \mathrm{dups} \} )$ } 
\State {\small view\_change = Distance($V_{curr}$,$V_{prev}$)}
\State {\small \emph{l} = \emph{l} - $\mid$ T $\mid$}    {\scriptsize   // remove user labeled pairs from budget}
\EndWhile \\ 
Return ${\mathbf{V_{curr}}}$
\end{algorithmic}
\end{algorithm}

\subsubsection{Subsequent Training Examples}\
\label{sec:cleaning:subsequent}

To improve the initial classifier, the active learning method selects
additional training examples for the user to label.  As described in
Section~\ref{sec:background}, active learning strives to select
examples that are most informative and thus have the highest
potential to help improve the classifier. It then learns a new
classifier on the expanded training data. 

As in the case of the initial classifier, we propose to take a
different approach and leverage view impact scores when selecting
additional training examples. We propose the following two approaches:



\textbf{View Impact Method (ViewImpact): } As in the case of learning the initial
classifier, this approach favors training examples where at least one
tuple has a high view impact score. Instead of breaking ties randomly,
however, we select those
samples that can help improve the current classifier the most. The tie
breaker is to select the samples with small {\it margin distance}, \ie the samples with the minimum absolute confidence
score from the classifier. 
If the margin distance is small, the classifier is less
confident. Therefore, the sample is better chosen for labeling as it
should help fine-tune the classifier. This tie-breaker is similar to
the uncertainty method, in which the examples that are closest
to the decision boundary are selected. In our case, however,
uncertainty is secondary to view impact.  This selection
algorithm corresponds to method $Select_{top}$(b,PS) in
Algorithm~\ref{EMDalgorithm} and it works as follows:

\begin{packed_item}
\item The function takes as input the \texttt{PairScores} data
  structure. For each subset of pairs $\{ ((t, u), s) \} \in$
  \texttt{PairScores} associated with the same original tuple $t$, the
  approach retains only the entry with the lowest margin distance,
  which it appends to a new data structure, \texttt{TopPairScores}.


\item The algorithm then selects $b$ pairs from \texttt{TopPairScores} using
weighted sampling where the view impact score $s$ serves as the weight.

\end{packed_item}


\textbf{Hybrid Method (Hybrid): } We also explore a second approach
that is a hybrid between the traditional method of selecting the most
informative training examples and the ViewImpact method.  The
hybrid approach computes the same \texttt{TopPairScores} structure as
the ViewImpact method above. However, it then assigns the following
hybrid weight to each pair of tuples in that set before selecting the
next batch of $b$ examples using weighted
sampling. \texttt{ClassifierUncertainty} measures the classifier
uncertainty as in state-of-the-art active learning methods for record
deduplication~\cite{mozafari2015scaling} (See
Section~\ref{sec:background}).

\begin{equation}
\mathrm{Hybrid Score~} = \alpha \mathrm{View\-Impact\-Score} + (1 - \alpha) \mathrm{Classifier\-Uncertainty}
\label{hybrid-eqn}
\end{equation}


\noindent Hence, in contrast to ViewImpact, which weighs samples based
on their view impact score alone, Hybrid can weigh both the
ClassifierUncertainty and ViewImpactScore with an adjustable relative weight,
$\alpha$.


Algorithm~\ref{EMDalgorithm}, lines 12 and following capture how the
above methods fit within the overall active learning part of the
cleaning process.

\subsection {Stopping Condition}
\label{sec:stop}
The state-of-the-art stopping condition for
deduplication~\cite{gokhale:2014corleone} is based on when the
classifier stops improving in accuracy. The idea is to check the
\emph{confidence}, or agreement among classifiers on a set of example
pairs, over a fixed window of time. Before active learning, a small
(3\%) random sample of pairs from the underlying data is set aside as
a holdoutset for evaluating classifier quality. As the classifier
learns from more informative examples, the confidence values will
increase. However, when there are few informative examples left to
learn from, the confidences level off.  For example,
in recent prior work in~\cite{gokhale:2014corleone}, when the confidence values have
stabilized within +/- $\epsilon = 0.01$, over a window size of
$n_{converge} = 20$ iterations, then the training stops. 


Our approach, in contrast, is to check the convergence of the view
quality by measuring the view sensitivity as defined in
Section~\ref{sec:sensitivity}. Intuitively, the view cleaning process
should stop when the view stops improving or when the user has
exhausted her labeling budget. A view stops changing as a result of
cleaning either because the data has been cleaned or because the view
is generally not sensitive to the remaining duplicate tuples. We say that
a view has \textit{converged}:

\defn{A view has converged if \\Distance($V_{curr},V_{prev}$) $\leq \epsilon$ for $n_{converged}$ iterations.  
}\edefn

The condition for stopping the process of cleaning a view is thus
based on the convergence, within some $\epsilon$, of the
\texttt{Distance} function computed between consecutive views during
cleaning.

%% file: experiments.tex
\section{Evaluation}
\label{sec:eval}

We evaluate our View Impact Cleaning approach compared with
state-of-the-art view-agnostic active-learning.


\begin{table}[t!]
\begin{scriptsize}
\begin{center}
\tabcolsep=0.05cm
\begin{tabular}{|l|l|c|c|c|c|}\hline
Dataset & Rows & Pairs & Cols & Dup & Dups \\ 
with blocking method & & & & Pairs & (\%) \\ \hline
Restaurants:(Fodor $\cup$ Zagat) & 864 &  7.4x$10$\textsuperscript{5} &  5  & 224 & 0.03  \\
 
block on view:SFrestaurants &148 & 2.1x$10$\textsuperscript{4} & 5 & 36 & 0.17 \\
block on view \& features:SFrestaurants & 148 & 384 &5 & 36 & 9.4\\ \hline
Products:(Amazon $\cup$ Google )& 4,589 & 2.1x$10$\textsuperscript{7} & 4 & 1,300 & 0.006\\

block on view:MfrProducts& 291 & 8.4x$10$\textsuperscript{4} & 4 & 162 & 0.19\\
block on view: \& features:MfrProducts & 291 & 6.9x$10$\textsuperscript{3}  &4 & 162 & 2.3 \\ \hline
\end{tabular}
\end{center}
\end{scriptsize}
\vspace{-.6cm}
\caption{Datasets used in evaluation. Table shows the cardinality, number of pairs, degree, number of duplicate (matching) pairs, and fraction of pairs that are duplicates for the base data and after view and feature blocking.} 
\vspace{-.6cm}
\label{dataset-details}
\end{table}

\noindent \textbf{Classifier.} We use a common choice for
classification, support vector machines, to build the classifier. Our
implementation uses libsvm~\cite{chang2011libsvm}. We use either
linear kernels or Gaussian kernels by tuning over the data and set the
weights for the positive and negative label classes to the reciprocals
of their respective cardinalities.


\noindent \textbf{Features.} Selecting the right features to give to a
machine learning algorithm (or \emph{feature engineering}) is a
well-known, challenging problem. Recent prior work leaves the feature
selection decision to an expert user~\cite{anderson2013brainwash}, or
in the case of Corleone~\cite{gokhale:2014corleone}, the system
randomly selects a subset of attributes as features. For the
applications that we target, we cannot require that users define
features and thus, similar to Corleone, must rely on an automated
approach. In our implementation, we use generic, type-based features,
but any automated feature-selection approach could be applied. 

\noindent{\bf Data.} We use two datasets (Table~\ref{dataset-details})
that have been extensively used in prior deduplication
work~\cite{gokhale:2014corleone}. The restaurants dataset is the union
of restaurant data collected from the culinary rating sites, Fodor and
Zagat. 533 tuples come from Fodor and 331 from Zagat for a
total of 864 tuples and 745,632 distinct pairs to
classify. Figure~\ref{fig:emd-views} shows example records.
Products, a more challenging dataset to deduplicate, combines
electronics products from Amazon (1,363 rows) and Google (3,226
rows)~\cite{amazon-google-products} with schema (name, description,
manufacturer, price).   An example record is: [\emph{`learning
  quickbooks 2007',`learning quickbooks
  2007',`intuit',38.99}]. There are more than 21
million tuple pairs in the union of these tables, among which
only 1,300 pairs refer to the same entities (0.006\% matches).

\begin{table*}[t!]
\begin{scriptsize}
\begin{sf} 
\begin{center}
\tabcolsep=0.11cm
\begin{tabular}{l|l}
View & View SQL and Description \\
(Card, \% Affected, Initial Dist.) \\
\hline

SFrestaurants (Top3) 
& {\tiny SELECT cuisine, COUNT(*)} 
 {\tiny FROM SFrestaurantsSelect*}
 {\tiny GROUP BY cuisine} 
 {\tiny ORDER BY COUNT(*) DESC LIMIT 3} \\ 
(3, 33\%, 0.44) & Top 3 restaurants in San Francisco 
 by type of cuisine \\
\hline

SFrestaurants (Select*)
& {\tiny SELECT *}
  {\tiny FROM restaurants}  
  {\tiny WHERE city = `SF'} \\ 
(148,12\%,0.31) & All restaurants in San Francisco \\
\hline

SFrestaurants (Count*)
&
{\tiny SELECT COUNT(*)}
{\tiny FROM SFrestaurantsSelect*} \\ 
(1,100\%,0.17) & Count of restaurants in San Francisco  \\
\hline

SFrestaurants (JoinAvgScore)
& {\tiny SELECT cuisine, AVG(score)}
{\tiny FROM SFrestaurantsSelect-scores} 
{\tiny GROUP BY cuisine}  \\
(29,31\%,0.13) & Restaurants by cuisine \& AVG inspection score 
from the San Francisco Health Department's 
restaurant inspection scores DB~\cite{SFhealthdept} \\
\hline

SFrestaurants (GroupByCuisine)
& {\tiny SELECT cuisine, COUNT(*)}
  {\tiny FROM SFrestaurantsSelect*} 
 {\tiny GROUP BY cuisine} \\ 
(29,31\%, 0.08) & A histogram-like view of restaurants
 in San Francisco grouped by cuisine \\
\hline

MfrProducts (Select*)
& {\tiny SELECT *}
{\tiny FROM products} 
{\tiny WHERE name LIKE `\%Apple\%' }
{\tiny OR name LIKE `\%Microsoft\%'} 
{\tiny OR name LIKE  `\%Adobe\%'} \\
(291,27\%, 0.47) & Products manufactured by 
Apple, Microsoft, or Adobe \\
\hline

MfrProducts (Count*)
& {\tiny SELECT COUNT(*)} 
{\tiny FROM MfrProductsSelect*} \\
(1,100\%, 0.30) & Count of products manufactured
by Apple, Microsoft, and Adobe \\
\hline

MfrProducts (PriceBins)
& {\tiny SELECT mfr, CASE} 
 {\tiny WHEN price $<$ 10 then 'Bin 1: [0,10)'}
 {\tiny WHEN price $<$100 then 'Bin 2: [10,100)'} 
 {\tiny WHEN price $<$ 1000 then 'Bin 3: [100,1000)'} 
 {\tiny ELSE 'Bin 4: 1000+'} \\
&  {\tiny END AS priceRange,}
 {\tiny COUNT(*) FROM MfrProductsSelect*}
 {\tiny GROUP BY mfr, priceRange} 
 {\tiny ORDER BY mfr ASC, priceRange ASC LIMIT 5} \\
(5,20\%,0.28) &For each manufacturer, tally the products 
 in various price ranges limited to the first 
 5 groupings \\
\hline

MfrProducts (Top3)
& {\tiny SELECT mfr, COUNT(*) as cnt}
{\tiny FROM MfrProductsSelect*} 
{\tiny GROUP BY mfr} 
{\tiny ORDER BY cnt DESC LIMIT 3} \\
(3,33\%,0.15) &Top 3 manufacturers  sorted on 
total count in descending order \\

\end{tabular}
\end{center}
\end{sf}
\end{scriptsize}
\vspace{-.5cm}
\caption{Views studied on restaurants and products: view sizes, fraction of rows impacted by duplicate entities, and initial view sensitivities to duplicates using $Distance(View_{dirty}$, $View_{clean}$) where $View_{clean}$ is the true clean view.} 
\vspace{-.3cm}
\label{all-view-sensitivity}
\end{table*}

\noindent {\bf Views.} We study \texttt{SELECT}, \texttt{PROJECT}, and
\texttt{AGGREGATE} (\eg \texttt{GROUP/ORDER BY LIMIT}) 
views. Queries containing joins have not been evaluated, but there is
no theoretical limitation to applying the View Impact Cleaning method
to such views. Recent prior work on deduplicating
views~\cite{altwaijry:2013query} only applies to simple, non-aggregate
\texttt{SELECT}/\texttt{PROJECT} views, while others such as
SampleClean~\cite{wang2014sample,krishnan37stale} are designed only
for aggregate queries without ordering nor top-k clauses.  We evaluate
our approaches over 9 views (Table~\ref{all-view-sensitivity}) that we
choose for the following reasons: (1) variety of impact that
individual duplicates have on the view and (2) variety of the overall
views' sensitivities to duplicates.




\noindent {\bf Blocking methods.}  
We use two types of blocking strategies to reduce the class imbalance
between positive and negative examples: view-based and feature-based
blocking. Table~\ref{dataset-details} shows how each type of blocking
increases the fraction of positive examples by an order of
magnitude. For the restaurants dataset, our views select restaurants
in San Francisco (shown as SFrestaurants). For the products dataset,
the views include products sold by Microsoft, Apple, and Adobe (shown
as MfrProducts). For feature-based blocking, for restaurants, we drop
pairs whose \texttt{jaccard} or \texttt{jaccard containment} match
scores on the \emph{name} and \emph{address} attributes respectively
are less than 0.2. For products, we block on \emph{price} and
\emph{name} when the \texttt{normalized Euclidean} distance on price is greater
than 0.54 and \texttt{jaccard} scores are less than 0.17 or
\texttt{jaccard containment} scores are less than 0.27 for \emph{name}. While we pick
these thresholds manually, the approach could be automated by
systematically dropping some percentage $P$ of least similar pairs along
each attribute. Please see~\cite{kmorton-thesis} for additional
details on the features computed and machine learning settings used.


For all experiments in this section, \textit{all approaches} (including
view-agnostic active learning) select pairs from the two views that
include \emph{both} blocking strategies, or SFrestaurants and
MfrProducts.

\noindent{\bf Experimental setup.}  We run the restaurants experiments 
20 times and the products experiments 100 times. For each
experiment, we create a randomly-selected holdoutset, which is not
used for training.  It serves to evaluate the quality of the
classifiers.  The holdoutset size is approximately equal to
half of the size of the initial unlabeled set.


\noindent{\bf Methods compared.} We apply two state-of-the-art active
learning methods, which we refer to as Uncertainty and
Entropy, as a baseline. For each of these methods, we use
an uncertainty or entropy measure to select the subsequent batches of
examples for active learning. Our implementation is based on the
description in~\cite{mozafari2015scaling}:
uncertainty~\cite{mozafari2015scaling,settles2010active} and
entropy~\cite{gokhale:2014corleone} scores are computed over 10
bootstraps that are sampled with replacement from the trainingset. The
examples are ranked by either their uncertainty or entropy scores and
selected by applying biased weighted sampling.  Both Uncertainty and
Entropy measure the disagreement of the classifiers over the
holdoutset example labels. Thus, the higher the uncertainty, the
stronger the disagreement, and the more informative the example
is to the learner.

\subsection{End-to-End Results}

\begin{figure*}[t!]
\begin{center}
    \includegraphics[width=0.8\textwidth]{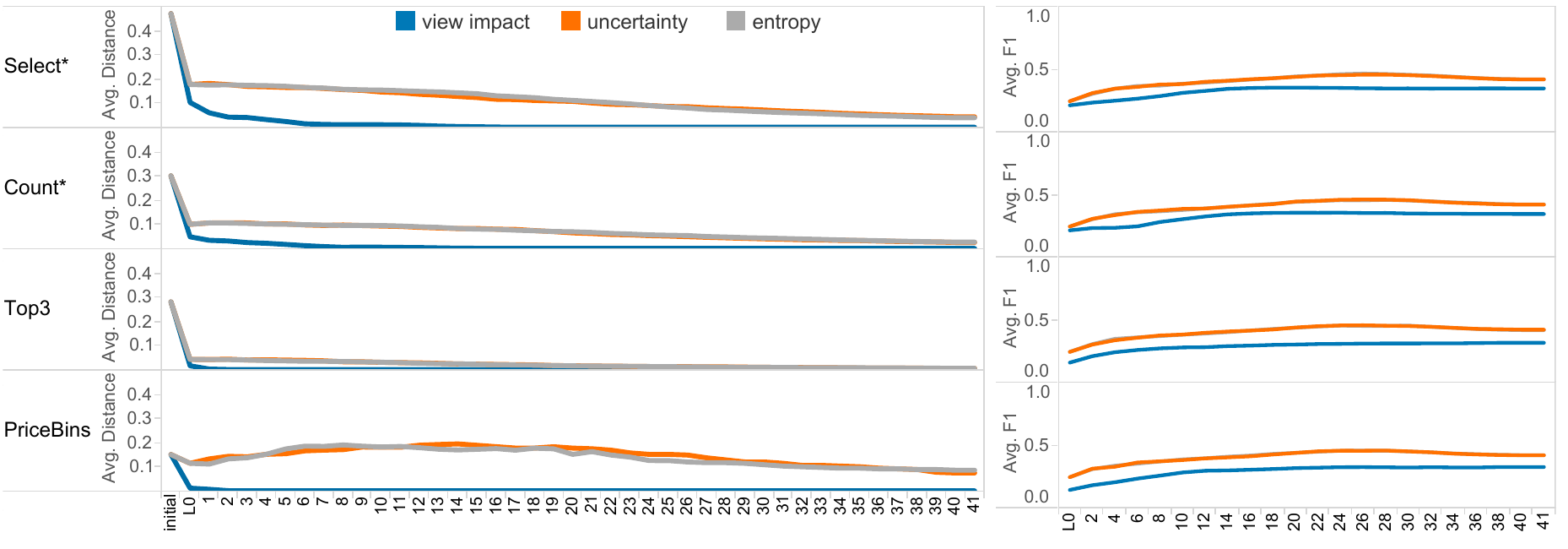}
    \vspace{-.7cm}
\end{center}
      \caption[All approaches cleaning views on the products dataset]{All product views are cleaned by View Impact (monotonically decreasing) in less than 18 batches with an $L_0$ size of 100 pairs and subsequent batch size of 20. View cleanliness (left) is shown with AVG $Distance(View_{curr},View_{clean})$ and (right) classifier accuracy with avg. F1. Entropy and Uncertainty have similar results for both cleaning ability and classifier accuracy: both require more than 42 batches to completely clean any of the product views and exhibit non-monotonic behavior when cleaning the \texttt{PriceBins} view.}
\vspace{-.5cm}
  \label{all-results-products}
\end{figure*}

\begin{figure}[t!]
\begin{center}
    \includegraphics[width=1.1\linewidth]{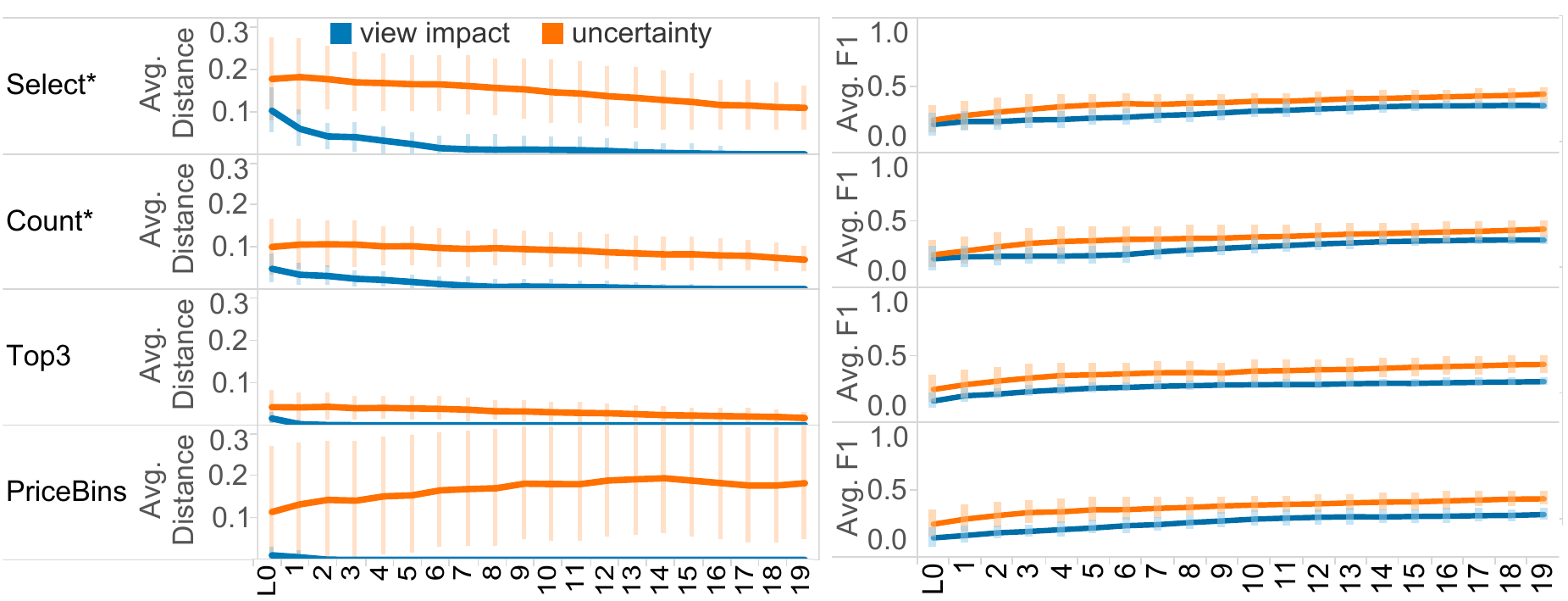}
    \vspace{-.9cm}
\end{center}
\caption{Results from Figure~\ref{all-results-products} zoomed in
on active learning part of cleaning and with error bars showing standard
deviations.}
\vspace{-.5cm}
  \label{main-results-products}
\end{figure}

\begin{figure}[t!]
  \centering
    \includegraphics[width=1.1\linewidth]{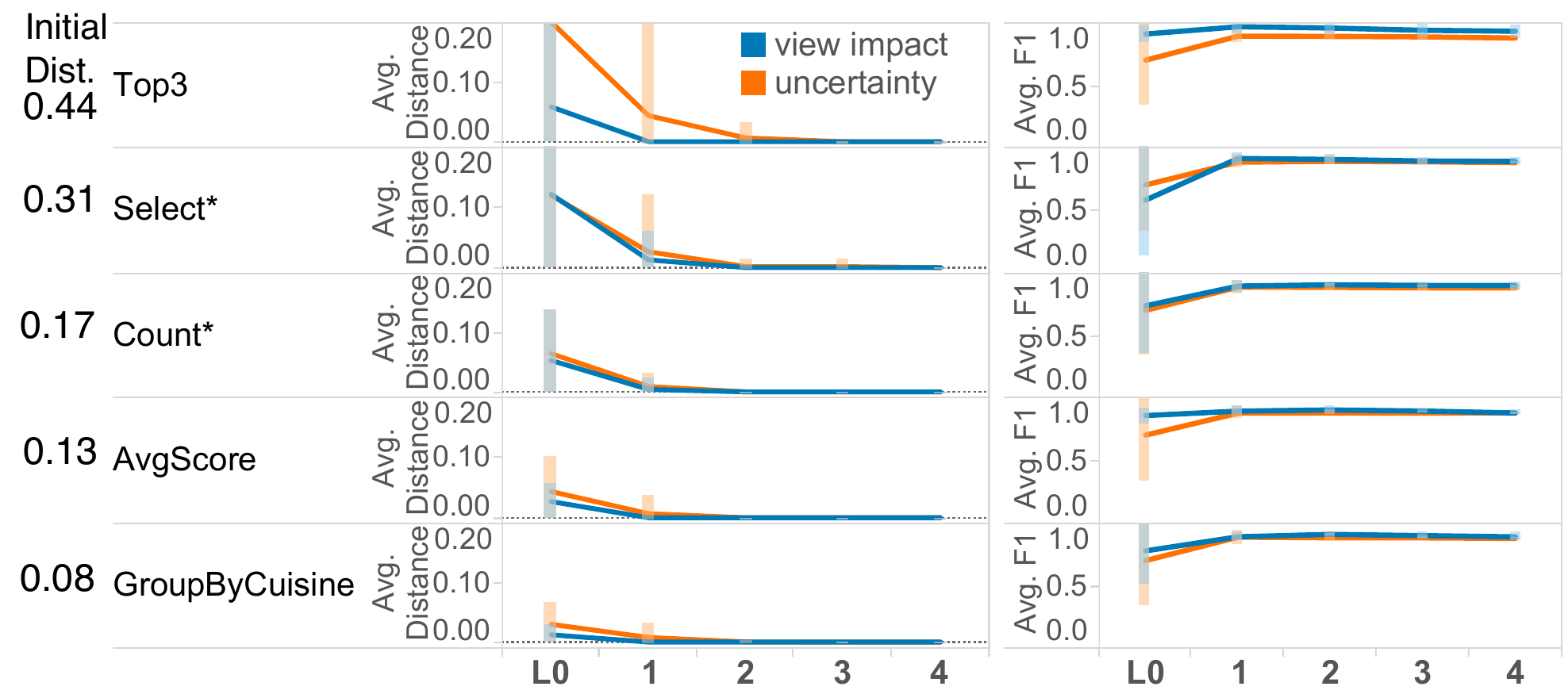}
    \vspace{-.7cm}
  \caption{All restaurant views are completely cleaned after three batches with view impact and four batches with related work. View cleanliness (left) is shown with avg. $Distance(View_{curr},View_{clean})$ +/- $\sigma$ and (right) classifier accuracy with avg. F1. Related work is Uncertainty. The initial classifier is trained on 13 pairs and subsequent batches are of size 20 pairs.}
  \vspace{-0.5cm}
  \label{main-results-restaurants}
\end{figure}

We first compare the overall ability of our approach, View Impact
Cleaning, and the two state-of-the-art active learning algorithms,
Entropy and Uncertainty, to clean views with a small number of user
labels.
Figures~\ref{all-results-products},~\ref{main-results-products},
and~\ref{main-results-restaurants} show \texttt{Distance}$(V_{curr},
V_{clean}$) before cleaning (value shown under ``Initial distance''),
after cleaning with the initial classifier (labeled $L_0$), and after
the maximum budget of user labels. Each point represents the average
of either 20 runs (Restaurant dataset) or 100 runs (Products dataset)
and the standard deviation, $\sigma$.  Since
Figure~\ref{all-results-products} shows that Uncertainty and Entropy
achieve similar cleaning results and classifier accuracies, we present
only Uncertainty in future graphs.


\noindent {\bf Products main result.} As shown in
Figures~\ref{all-results-products}
and~\ref{main-results-products}(left), all product views are
completely cleaned by View Impact in fewer than 18 batches, or 440
labels (including the initial $L_0$ batch of size 100 and subsequent
batches of size 20). \texttt{Top3} and \texttt{PriceBins} are cleaned
in only three and four batches, respectively (140 to 160
labels). These views are cleaned faster with View Impact than the
\texttt{Select*} and \texttt{Count*} views because a smaller number
of tuples impacts these views. These are the only tuples with non-zero
view impact scores and View Impact biases the selection of tuple-pairs
for the user to label toward these tuples.  In contrast, all tuples
impact \texttt{Select*} and \texttt{Count*} views and do so equally,
leading to a longer cleaning process.  Most importantly, for all
views, the View Impact Cleaning approach yields rapid improvements in
view quality early on in the cleaning process. For the
\texttt{Select*} view, our approach cuts the distance to the clean
view by 4X after the initial classifier (from 0.47 to 0.11). The first
subsequent batch cuts the distance by another 50\%. In contrast, the
Classifier Uncertainty and Entropy methods are unable to completely
clean any views in 40 batches.


Figures~\ref{all-results-products}
and~\ref{main-results-products}(right) show the classifier accuracy
(F1score) achieved by all methods. It is difficult to build a
quality classifier for the products dataset as evidenced by the low
average F1 scores for all methods. The products dataset contains many
data quality problems including missing and wrong values, which
complicates feature selection. For example, \emph{name} values were
inconsistent even for matching pairs.  We thus used this attribute for
blocking but not for learning. We observe, however, that as expected
the View Impact Cleaning method yields, on average, a classifier with
a lower F1 score than the Uncertainty or Entropy methods. This method
focuses on the quality of the view rather than the quality of the
classifier itself.






Interestingly, all views are cleaned monotonically with the View
Impact Cleaning approach, while some aggregate views such as
\texttt{PriceBins} exhibit non-monotonic behavior for the other
methods. We see this undesirable behavior with Uncertainty and Entropy
because they focus on selecting examples that improve the classifier's
quality and not the view.  Since the classifier's accuracy is low, it
is unable to correctly label the pairs that impact the view. The View
Impact Cleaning approach, in contrast, favors as training examples
those pairs that have a high impact on the view. Since these labels
are not the most informative, the classifier it learns is not as good
as Uncertainty, but these labels are useful for cleaning the view.

\noindent {\bf Restaurants main result.} 
Figure~\ref{main-results-restaurants}(right) shows that all approaches
exhibit higher average F1 scores on the holdoutset for restaurants
than products, which suggests that duplicates in this dataset are much
easier to classify.  We thus expect that the results for cleaning
with all approaches should be similar. We observe, in
Figure~\ref{main-results-restaurants}(left), that the View Impact
Cleaning method cleans all restaurant views in three batches, while
Uncertainty needs four batches.  Assuming that the initial classifier
is learned over a batch of 13 pairs and subsequent batches contain 20
pairs, view impact can clean all views one batch faster than
Uncertainty.  Furthermore, for the \texttt{Top3} view, view impact
only requested two batches (33 labels), while Uncertainty required two
additional batches of 20. These results indicate that even when a good
classifier can be learned with a small number of labeled examples, our
technique does not hurt the quality of the view compared with
Uncertainty.


\subsection{Learning an initial classifier}
\label{initial-classifier}
We now study the individual components of the View Impact Cleaning
approach.  The first component of the approach is the selection of the
initial training examples (C.F. Section~\ref{sec:cleaning:initial}). The selection
occurs after both view-based and feature-based blocking.

We measure the quality of the view obtained after cleaning using the
initial classifier learned with View Impact Cleaning. We compare the
results to cleaning when using a classifier learned on a strictly
random sample of the data taken also after both view-based and
feature-based blocking. As discussed in
Section~\ref{sec:cleaning:initial} and as shown in
Table~\ref{dataset-details}, because the number of positive examples
is extremely small compared with the number of negative examples, an
initial classifier learned on a random data sample may have no
positive examples to learn from. We thus also compare with a third
approach that biases the selection of the training examples to select
a larger fraction of positive examples. We call this last method
\textit{per-feature round-robin}. This approach sorts the tuple pairs
by decreasing value of each of their features. It creates as many
sorted lists as there are features and each pair appears once in each
list. It then uses weighted sampling to select the pairs using the
rank in the lists as weight. 

  

\begin{figure}[t!]
\begin{small}
\begin{scriptsize} 
\begin{center}
\includegraphics[width=1\linewidth]{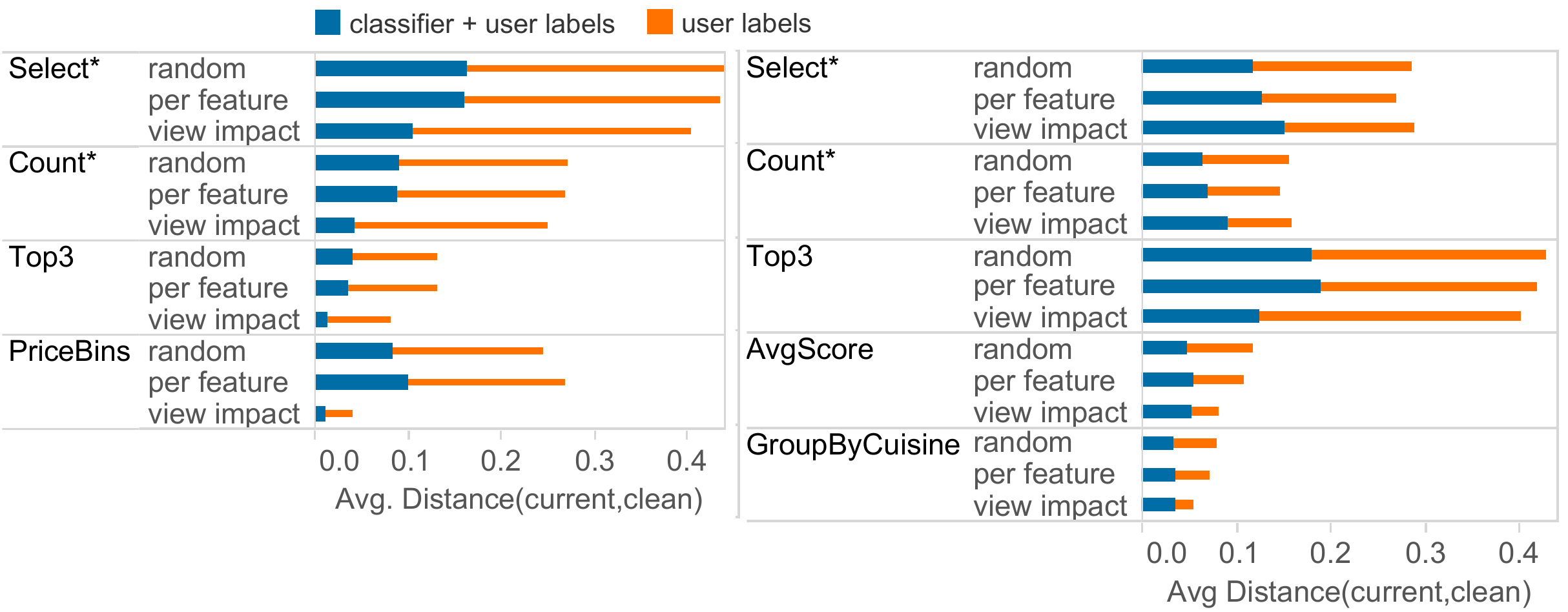}
\tabcolsep=0.05cm
\begin{tabular}{| l | l | c | c| c|} \hline
 &  & \multicolumn{3}{c|}{Average F1} \\ \hline
Dataset & View &  Random & Per feature & View impact  \\ \hline
Products & Select* & 29\%&32\% &27\% \\
Products & Count* & 29\%&32\% &28\% \\
Products & Top3 & 29\%& 32\% & 28\% \\
Products & PriceBins & 29\%& 32\%& 17\% \\ \hline
Restaurants & Select*& 84\%& 84\% & 71\%  \\
Restaurants& Count* & 84\%& 84\% & 65\%  \\
Restaurants& Top3 & 84\%& 84\% & 94\%  \\
Restaurants& AvgScore & 84\%& 84\% & 73\%  \\
Restaurants& GroupByCuisine & 84\%& 84\% & 74\% \\ \hline
\end{tabular}
\end{center}
\end{scriptsize}
\end{small}
\vspace{-.6cm}
\captionof{figure}{Impact of biasing the initial classifier on avg. Distance to the true clean view for products (left) and restaurants (right) and avg. F1 (table below). Initial $L_0$ batch size is 100 for products and 13 for restaurants.}
\vspace{-.5cm}
\label{fig:all-L0-biasing}
\end{figure}

\noindent {\bf Products.} Figure~\ref{fig:all-L0-biasing} (left) shows
the results for the four views over the Products dataset. As the
figure shows, the View Impact Cleaning method yields the cleanest
views after this initial cleaning step.  Because View Impact Cleaning
focuses on labeling and cleaning pairs with tuples that have
high-impact on the view, an important question that arises is whether
a classifier is at all useful or whether all the benefits come from
the user labels. The figure also shows the quality of the view if we
clean it using only the user labels. As the figure shows, with all
three methods, building and using a classifier is critical to cleaning
the view. Interestingly, the classifiers help to clean the view
even though their average F1 accuracies are low for all sampling approaches (see Figure~\ref{fig:all-L0-biasing}'s table).
This result implies that, for the purpose of quickly cleaning a view, it is not necessary to learn a high-quality classifier; rather it is more important to have the user resolve the most impactful tuples first, and train a classifier using these biased labels.  \\
\noindent {\bf Restaurants.} We see in
Figure~\ref{fig:all-L0-biasing}(right) that all sampling strategies
have similar behaviors when cleaning the views for a small $L_0$ batch
of 13 example pairs, which corresponds approximately to 3\% of the
data (a commonly
used initial trainingset size~\cite{gokhale:2014corleone}). Since tuple-pairs in the restaurants dataset are easier to classify, View Impact Cleaning does not have as much of an advantage as before. However, for three out of five views, View Impact Cleaning still produces a cleaner view than the other approaches. \\

\subsection{Tuning the parameter settings}

In this section, we study the effect of tuning various settings for
the View Impact Cleaning and Uncertainty approaches.  We first present
the impact of weighting the two cleaning approaches using the
$\alpha$ parameter in Equation~\ref{hybrid-eqn} for the Hybrid
method that combines View Impact Cleaning with ClassifierUncertainty.
We also study the impact of varying the batch sizes.



\subsubsection{$\alpha$ values}


Since a good quality classifier can help save the user in cleaning
effort, we study the effect of the weighting factor $\alpha$ in the
Hybrid method described in Section~\ref{sec:cleaning:subsequent}.
Recall that this method selects pairs for labeling by assigning them
the following weight: $\alpha*ViewImpactScore + (1 -
\alpha)*ClassifierUncertainty$. We consider the two extremes:
prioritizing pairs that will improve the classifier ($\alpha=0$) and
prioritizing pairs that impact the view ($\alpha=1$).  We also
consider the hybrid method with $\alpha=0.5$.  Our analysis focuses on
the products dataset, since the quality of its classifiers was much
lower than those for restaurants. Furthermore, we zoom in to the
details for two views, which exhibit very different sensitivities to
duplicates. Other views showed similar trends.

\begin{figure}[t!]
\begin{scriptsize}
\begin{center}
\includegraphics[width=0.9\linewidth]{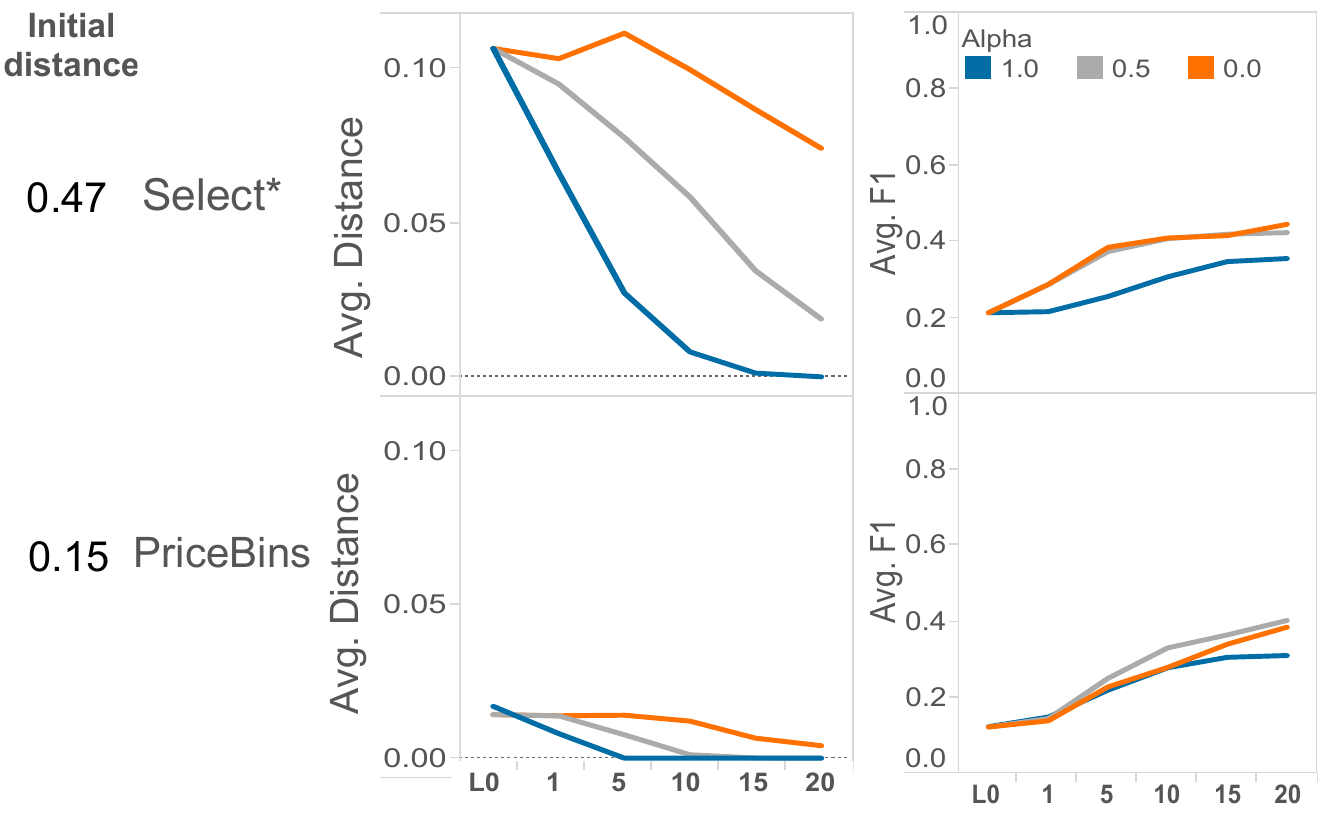}
\begin{tabular}{| l | c | c | c | c | c | c|} \hline
   \multicolumn{7}{|c|}{Stdev of Distance} \\ 
 & \multicolumn{2}{c|}{$\alpha=0.0$} & \multicolumn{2}{c|}{$\alpha=0.5$} &  \multicolumn{2}{c|}{$\alpha=1.0$} \\ 
 View & $\sigma_{L_0}$ & $\sigma_{20}$&$\sigma_{L_0}$ & $\sigma_{20}$&$\sigma_{L_0}$ & $\sigma_{20}$ \\\hline
Select* & 0.05 & 0.04 & 0.02 &0.02 & 0.05 & 0\\
PriceBins & 0.02 & 0.007 & 0.02 & 0.01 & 0.02 & 0 \\ \hline
\end{tabular}
\vspace{-.3cm}
\caption{Impact of $\alpha$ on avg. $Distance(V_{curr}$,$V_{clean}$) and avg. F1 for product views.
Initial $L_0$ has 100 pairs.}
\vspace{-0.5cm}
\label{fig:product-alphas}
\end{center}
\end{scriptsize}
\end{figure}

{\bf \noindent View with higher relative sensitivity to dups.}
Figure~\ref{fig:product-alphas} shows the result for the
\texttt{Select*} view, which has the highest initial sensitivity to
duplicates, 0.47.  All initial classifiers are trained on 100 example
pairs with View Impact Cleaning. The choice of $\alpha$ affects only
subsequent batches. As the figure shows, the View Impact Cleaning
approach (\ie $\alpha$ = 1.0) is still able to make more progress
cleaning than both the hybrid ($\alpha$ = 0.5) and
ClassifierUncertainty ($\alpha$ = 0.0), despite having consistently
lower overall classifier accuracy. In fact, View Impact Cleaning is
the only technique that completely cleans this view within the budget
of 500 labels (20 batches). This result suggests that heavily biasing
the selection strategy toward the most impactful pairs is better for
cleaning views that are highly sensitive to duplicates and defined
over a dataset for which it is difficult to build a high quality 
classifier.




{\bf \noindent View with lower relative sensitivity to dups.}
Figure~\ref{fig:product-alphas} shows the results for the
\texttt{PriceBins} view, which is close to half as sensitive to
duplicates as the \texttt{Select*} view. Once again, View Impact
Cleaning outperforms the other approaches. It is able to completely
deduplicate the \texttt{PriceBins} view by batch 5; neither of the
other two approaches could to do so by the end of the 500 label budget (20 batches).





\subsubsection{Batch size}

\begin{figure}[t!]
\begin{center}
\includegraphics[width=0.8\linewidth]{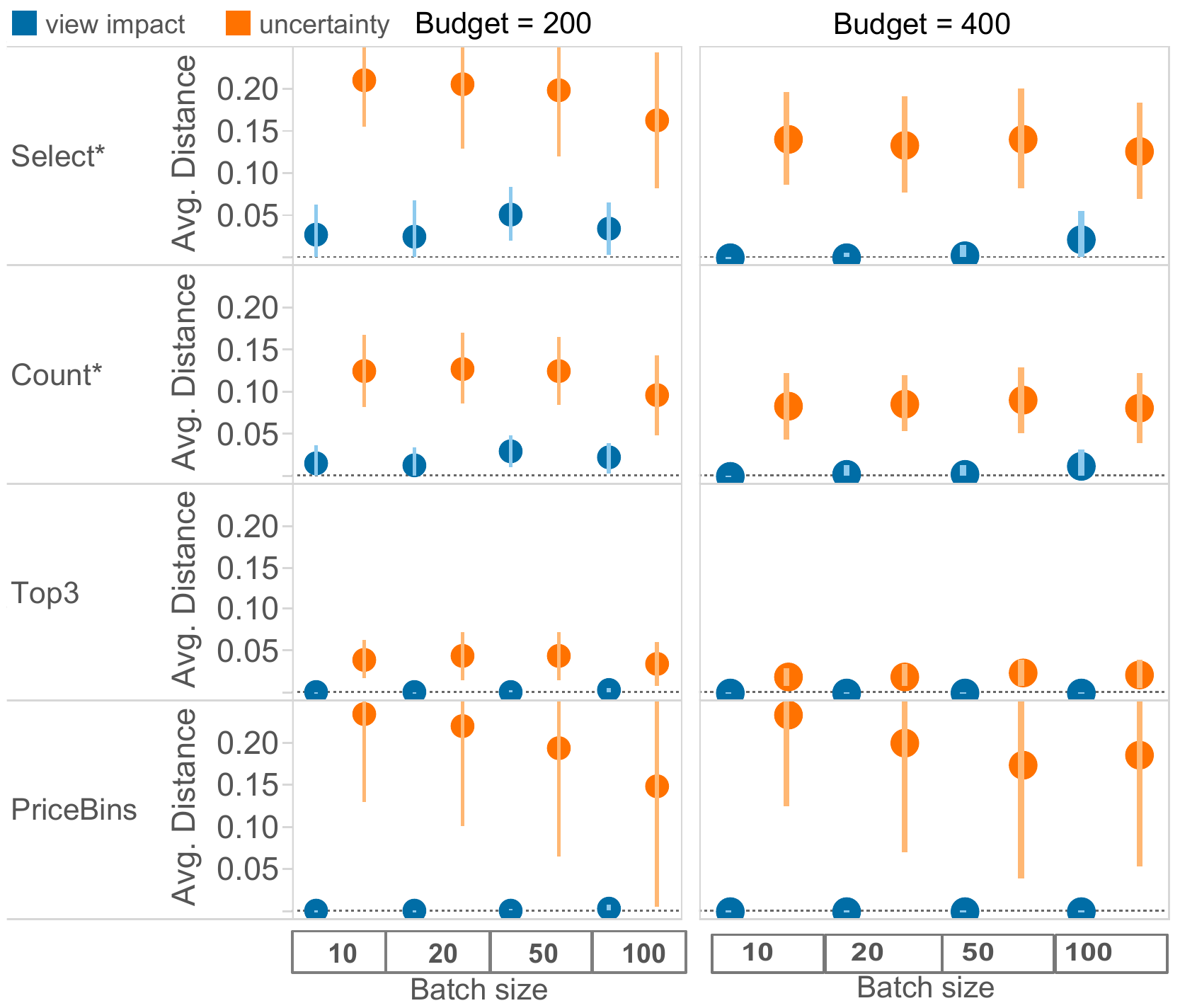}
\vspace{-.4cm}
\caption{For all product views, we see the impact of batch size on average $Distance(V_{curr}$,$V_{clean}$) +/- $\sigma$ with a budget of 200 (left) and budget of 400 (right).}
\vspace{-0.6cm}
\label{fig:all-batch-sizes}
\end{center}
\end{figure}

We study the effect of cleaning views with different batch sizes (10,
20, 50, and 100 example pairs) and budgets (400 and 200) with View
Impact Cleaning and Uncertainty. We show the result for products in
Figure~\ref{fig:all-batch-sizes}.  We observed similar results for
restaurants. Overall, the batch size does not significantly influence
the results. For all configurations, View Impact Cleaning is able to
clean more than Uncertainty on average.  Additionally, the variance
for Uncertainty is much higher than for View Impact Cleaning. This
result suggests that View Impact Cleaning is a more stable approach to
deduplication and that it is not sensitive to the batch size.

\subsection{Runtime and scalability}


\noindent{\bf View Impact Cleaning complexity.} There are two primary sources of computational complexity for the View Impact Cleaning algorithm. First, computing the feature vectors for all pairs is O($n^2$), where $n$ is the input dataset size. Second, computing the Distance as EMD in View Impact Scores (from Algorithm 1) takes worst-case O($n\times m^2$) because the EMD has O($m^2$) complexity where $m$ is the size of the view~\cite{ling2007efficient} and can be called (worst-case) $n$ times if $|Provenance(V(R))| = |R|= n$. Since computing the EMD grows quadratically with the view size, this approach works best with small views. Interestingly, a recent study of visualizations/views created on Tableau Public and \me~\cite{morton2014public} showed that 53\% of views have fewer than 1,000 rows. We discuss the empirical findings next.

\noindent{\bf Empirical runtimes.} We run View Impact Cleaning on a desktop machine with dual 2.4 GHz quad-core Intel Xeon processors and 11GB of memory. We use SQLite as our backend database to compute the feature vector table. We present the detailed measurement of runtimes for our approach on the view, \texttt{Select*} from products, as this view is the largest of all from Table~\ref{all-view-sensitivity} (with 291 rows) and takes the most time. We assume the same experiment settings as prior experiments on this view, where the initial $L_0$ batch has 100 pairs and subsequent batches have each 20 pairs. We time each of the key steps as follows:
(1) Compute view impact scores for all tuples: three minutes, 
(2) Compute feature vector with four features with view blocking and feature
blocking: three minutes (without feature blocking the time is 53 minutes)
(3) Pick examples to label per batch: under one second,
(4) Learn a new classifier per batch: under one second,
(5) Labels all pairs as either duplicates or not per batch: under three seconds.


As expected, steps (1) and (2) are the only steps that take a
significant amount of time.  To help with overall interactivity, these
steps can be done as a background process while the user first
explores the data. Interestingly, these two steps only need to
be performed once before the cleaning process begins. Over the course of cleaning a view, the tuple view impact scores tend to not change.



\begin{figure}[t!]
\begin{center}
\includegraphics[width=0.8\linewidth]{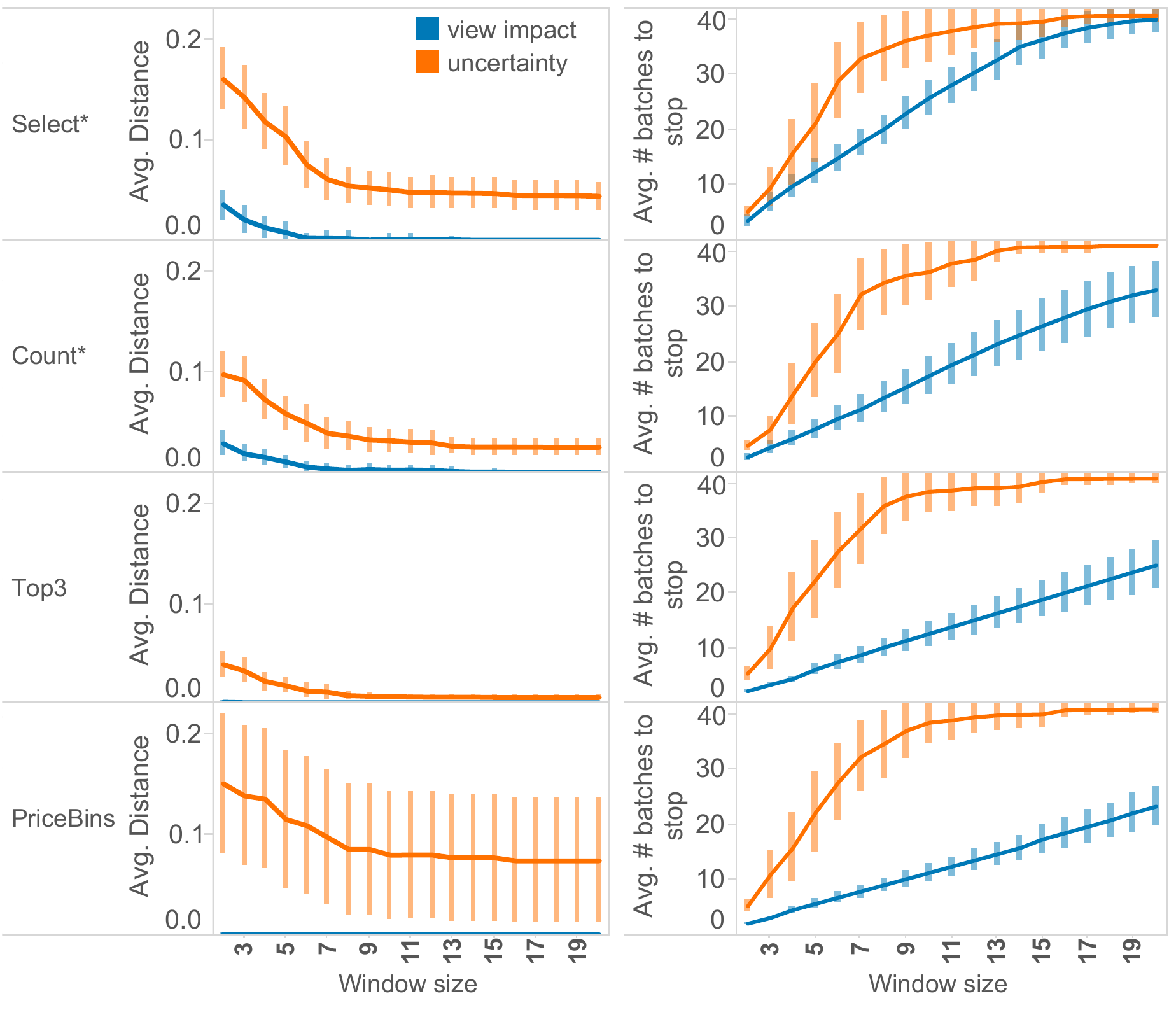}
\vspace{-.4cm}
\caption{For all views, View Impact Cleaning converges to a perfect clean view at a window size of 16, but Uncertainty does not converge for any window size (left is avg. Distance($V_{dirty}$,$V_{clean}$)+/- $\sigma$). Smaller window sizes can conserve labeling effort (right is avg. \#of batches), but may result in a less than perfect clean view.}
\vspace{-0.7cm}
\label{fig:products-stopping-condition}
\end{center}
\end{figure}

\subsection{Stopping condition}

Recall that in practice $V_{clean}$ is not known. We thus do not know
exactly when to stop cleaning. The heuristic used is to stop after
little to no progress has been made for some interval of time.  All we
can do is show empirically that this heuristic is effective.  The
Entropy method does this based on the stability of the confidence
values of the classifier. Since the Uncertainty approach does not
specify when learning can stop, we apply the same approach as used in
the Entropy work to monitor the stability of the uncertainty values of
the classifier. The View Impact Cleaning approach has a more natural
and direct way to measure ``little change" based on the
$Distance(V_{curr}$,$V_{prev}$). The idea is to stop cleaning once we
observe that the distances computed between the current view,
$V_{curr}$, and the view cleaned from the previous iteration,
$V_{prev}$, have plateaued (within +/- $epsilon = 0.01$) over a window
of size $n_{converge}$ batches. For the product views, shown in
Figure~\ref{fig:products-stopping-condition}, we evaluate the impact
of the window size on the convergence to the true clean view.  The
figure shows the distance values when cleaning stops (left) and the
corresponding labeling effort (right).


Given a much larger budget this time (900 labels, 41 batches total) and using the same $n_{converge}$ = window size = 20 batches and $\epsilon$ as reported in the Entropy evaluation section, Uncertainty still fails to converge to the true clean view for all product views. The result on the right indicates that the Uncertainty approach is unstable for a long time: all window sizes require many more batches to stop for Uncertainty than View Impact Cleaning. This result suggests that the Uncertainty classifier learned is not stable enough to stop given even a large labeling budget of 900. For View Impact Cleaning, we see that a window size of 16 achieves the objective of converging to the true clean view (\ie Distance = 0) for all views. However, each view requires between 18 to 37 batches of labels given this window size. While this result is consistent with the time period in which the \texttt{Select*} and \texttt{Count*} views actually converge to the true clean view (see Figure~\ref{main-results-products}), the \texttt{Top3} and \texttt{PriceBins} views require significantly less cleaning effort (only three to four batches).  If the user is willing to trade off cleaning quality for effort, a window size of 7 would be an appropriate compromise, as half of the views are completely cleaned and the other half have a small average Distance, 0.002, from the true clean view (99\% clean).


\subsection{Multi-view deduplication} 

A dashboard is a collection of related visualizations or views typically over a common dataset.  In this section, we study the performance of two techniques for data cleaning in the context of such dashboards.


\begin{figure}[t!]
\begin{center}
\includegraphics[width=1\linewidth]{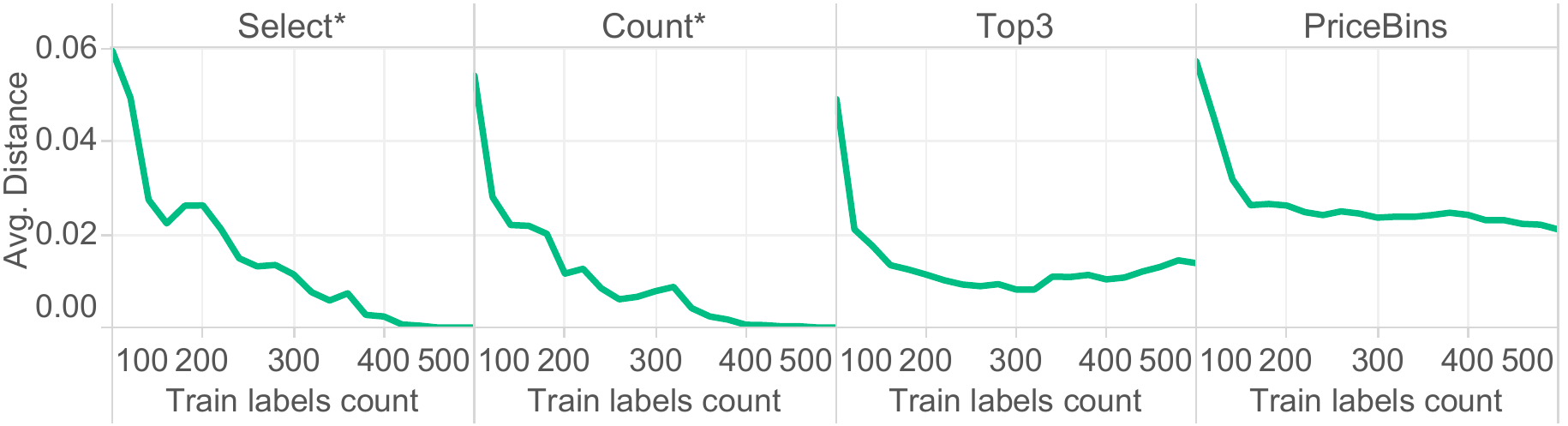}
\vspace{-.6cm}
\caption[Impact of fully cleaning each view on cleaning all other views for products.]{Effect of fully cleaning one view on cleaning the other views in products given a budget of 500 labels (batches = 20 pairs): resolving the duplicates in the \texttt{Select*} view (far left) helps clean all the other views the fastest in 460 labels. However, the views are not cleaned monotonically.}
\vspace{-0.4cm}
\label{fig:multi-view1}
\end{center}
\end{figure}

\begin{figure}[t!]
\begin{center}
\includegraphics[width=0.7\linewidth]{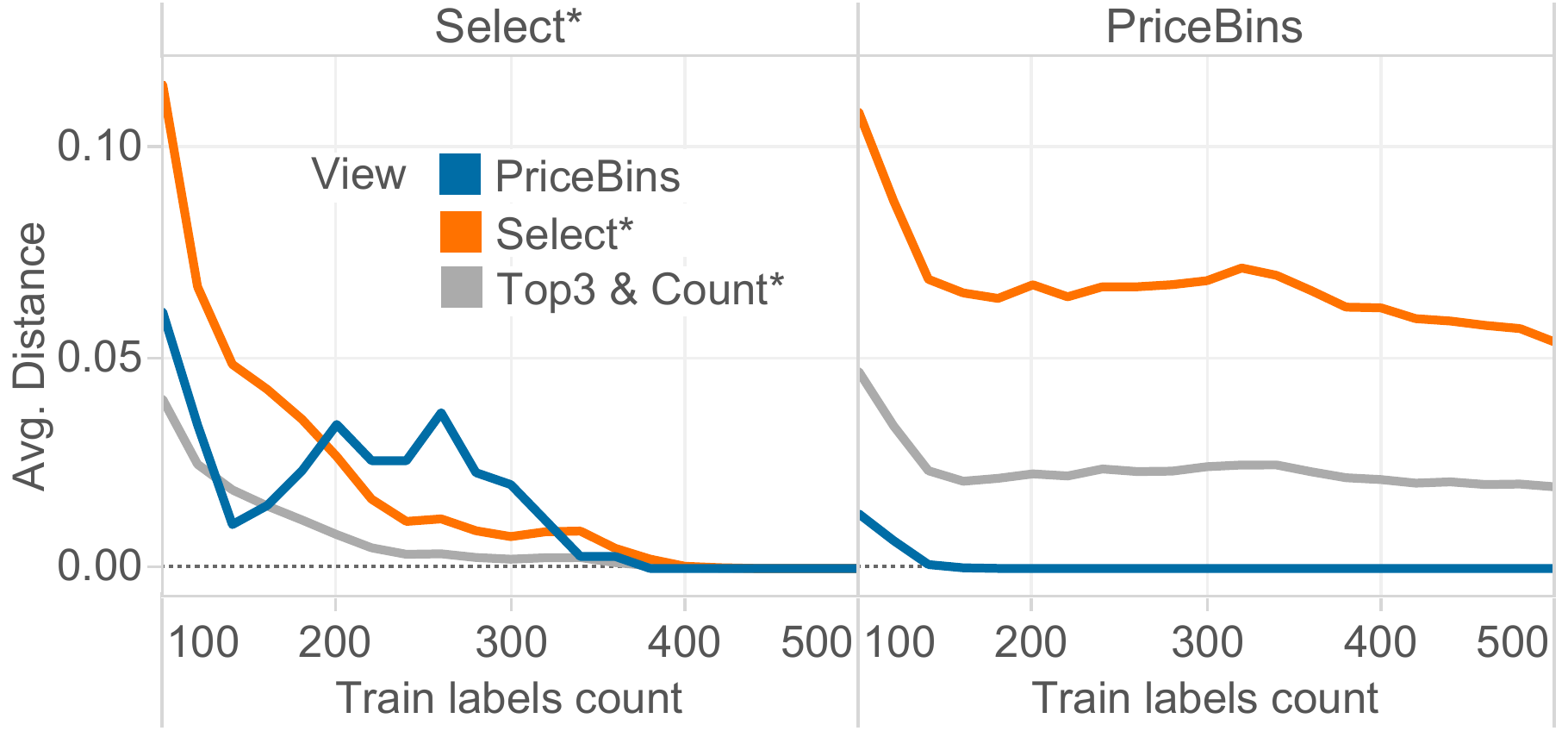}
\vspace{-.4cm}
\caption[Impact of cleaning the most sensitive view, \texttt{Select*}, and less sensitive view, \texttt{PriceBins}, on cleaning the other product views.]{The most sensitive view first approach (left, \ie cleaning \texttt{Select*}) monotonically cleans the other product views except for \texttt{PriceBins} and does so with only 460 labels. All batches contain 20 pairs.}
\vspace{-0.6cm}
\label{fig:multi-view2}
\end{center}
\end{figure}


\noindent{\bf (1) Fully clean one view at-a-time.} We first study how
much cleaning one view in a dashboard can help to clean the other
views. Figure~\ref{fig:multi-view1} shows the average distance,
$Distance(V_{curr}$,$V_{clean}$), across all four views for products as
we clean one of the four views only. As the figure shows, when
cleaning one of the two views with the greatest sensitivity to
duplicates, \texttt{Select*} and \texttt{Count*}, the most progress
can be made on simultaneously cleaning the other views:
\texttt{Select*} cleans all other views in 460 labels and
\texttt{Count*} cleans them in 480 labels. In these views, all tuples
have the same view impact scores and the cleaning process treats them
all in the same way helping to clean all views the fastest.
Interestingly, as shown in Figure~\ref{fig:multi-view2}, deduplicating
the \texttt{Select*} view using View Impact Cleaning causes temporary,
non-monotonic behavior in one view, \texttt{PriceBins}, which is
possible given that the quality of the classifiers is low and
subsequently learned classifiers may change how they classify
 the most impactful tuples for the \texttt{PriceBins} view.


Thus, cleaning one view helps to make progress on other
views. However, in the context of cleaning an entire dashboard, the
at-a-time method must be done with careful attention to the order in
which the views are cleaned.




\begin{figure}[t!]
\begin{center}
\includegraphics[width=0.7\linewidth]{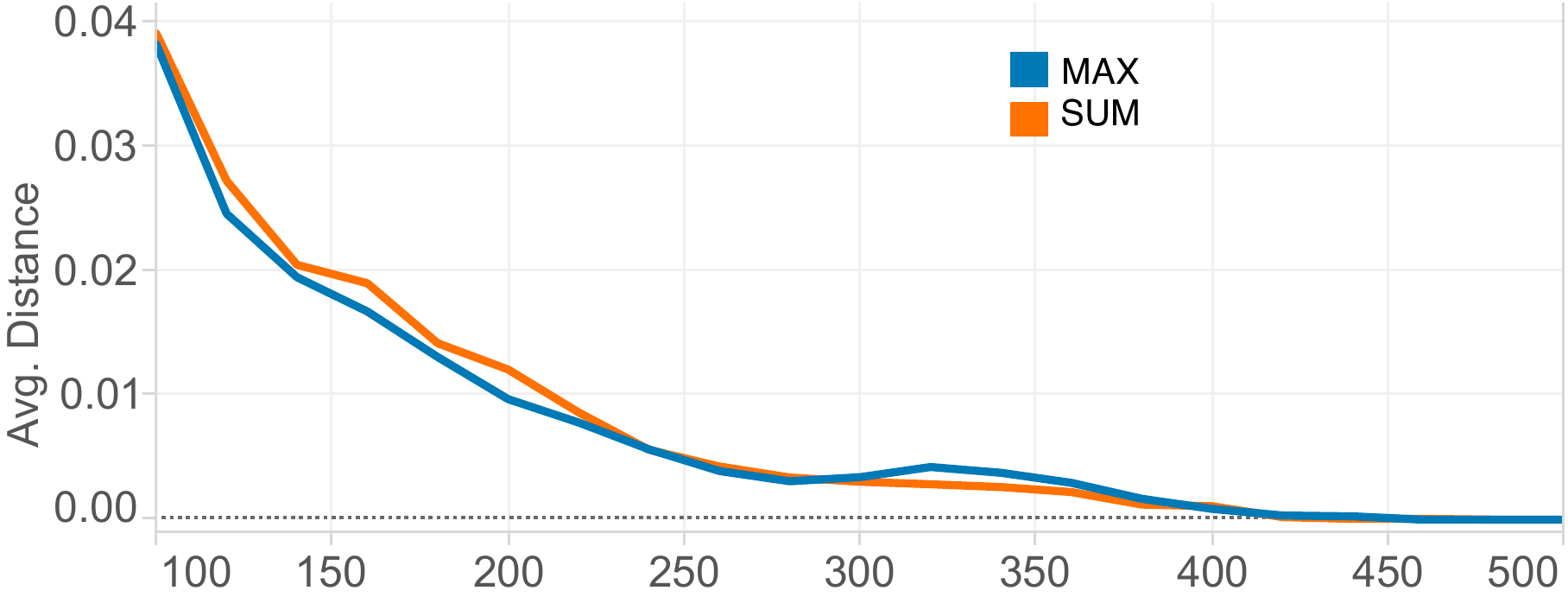}
\vspace{-.4cm}
\caption[Clean by max vs sum across all product views.]{MAX cleans all views simultaneously in 460 labels and SUM cleans in 480 labels (one batch later). All batches contain 20 pairs.}
\vspace{-0.6cm}
\label{fig:max-sum}
\end{center}
\end{figure}

\noindent{\bf (2) Clean across all views simultaneously using an aggregate measure of sensitivity across all views, MAX and SUM}.
We also evaluate the performance of cleaning a dashboard of visualizations. In this approach, the View Impact score for each tuple
in the base relation is either the max or the sum of its impact across all the views. Figure~\ref{fig:max-sum} shows the results. The results
are similar when using either MAX or SUM and the total number of labels required to clean all views is the same as cleaning
just \texttt{Select*} or \texttt{Count*}. However, the curves for both MAX and SUM are smoother than when cleaning only \texttt{Select*} (Figure~\ref{fig:multi-view1}). This approach has thus the double benefit of yielding more stable results across batches and avoiding the problem of
selecting which view to clean first.


\noindent{\bf Application to multi-view cleaning}. Beyond cleaning visualization dashboards, the results in this section show that if a user starts with one visualization, the effort spent cleaning that visualization will help speed up the cleaning of subsequent visualizations, even though View Impact Cleaning is a view-driven cleaning method. 


%% file: related.tex
\section{Related Work}
\label{s:related} 


Deduplication has a long history in the literature (see~\cite{halevy2012principles, getoor:2012entity}). The state-of-the-art deduplication approaches that are closely related to this work fall into three categories:

\noindent \textbf{Active learning.} The active learning systems from the literature focus on heuristics that select the minimum number of examples needed to learn a high quality classifier~\cite{beygelzimer2009importance} with the goal of cleaning an entire dataset at-a-time~\cite{arasu:2010active, Bellare:2012} and not a subset as in this work. Furthermore, other systems combine feedback from a set of crowd workers (who may provide incorrect or conflicting labels) with the goal of limiting the number of unnecessary label requests for resolving duplicates~\cite{wang:2012crowder, mozafari2015scaling, gokhale:2014corleone, vesdapunt2014crowdsourcing}. All of these systems (except Corleone~\cite{gokhale:2014corleone}) require a developer/expert to manage the common learning tasks such as writing the blocking rules, and creating training data for the matcher. Corelone pushes this expert work to the crowd.  In addition, Corleone extends common active learning methods by 1) applying biased sampling for the initial set, and 2) enforcing stopping conditions using an observation set.  Unlike previous work, we use View Impact Cleaning for sampling the initial set, selecting additional training examples, as well as designing the stopping condition. The other state-of-the-art crowd-based active learning algorithms in~\cite{mozafari2015scaling} use bootstrap~\cite{efron1994introduction} to estimate the classifier's uncertainty in its predictions of labels. ~\cite{vesdapunt2014crowdsourcing} applies a machine learned model that clusters similar records together (based on an associated probability of being a match).  However, all of these related systems often require the user to provide thousands of labels to clean entire datasets.  Our work, in contrast, saves the user's cleaning effort by focusing on resolving the data that has the greatest impact on the view.\\ 
\noindent \textbf{Passive learning.} A variety of techniques have been proposed for deduplication~\cite{gruenheid:2014incremental,welch:2012fast,whang:2014incremental}. Those works most relevant to us are learning-based techniques that train a classifier over a batch of labeled pairs of examples~\cite{kopcke2008training,bilgic2006d,kang2007geoddupe,whang2013question}. Many of these approaches try to reduce the label complexity by applying various feature-based similarity matchers and then sampling from the pairs that are likely to be matches or are the most \emph{informative}. However, we showed in Section~\ref{initial-classifier} that such biasing techniques are insufficient in completely cleaning any of the views in one shot. 

\noindent \textbf{Clustering.} Several deduplication approaches consider the setting where each tuple can match multiple other tuples~\cite{altwaijry:2013query,bhattacharya2007query,wang2013leveraging,vesdapunt2014crowdsourcing,wang2015crowd}. They either leverage the transitive property of the match relation~\cite{altwaijry:2013query,wang2013leveraging,vesdapunt2014crowdsourcing} or correlation clustering~\cite{bhattacharya2007query,wang2015crowd} to infer matching and non-matching pairs based on previously labeled pairs and reduce the labeling effort by users. These approaches are complementary to ours and could be added to our method to further speed up view cleaning.

%% file: conclusion.tex
\section{Conclusions}
\label{s:conclusion} 
We proposed an active learning algorithm for deduplicating records in an exploratory visual analytic system, which strives to produce the cleanest view possible within a limited budget. Our key idea is to consider the impact that individual tuples have on a visualization and to monitor how the view changes during cleaning. We demonstrated over a set of nine views that our approach produces significantly cleaner views for small labeling budgets than state-of-the-art alternatives and that it also stops the cleaning process after requesting fewer labels.

%% file: acks.tex
\section{Acknowledgements}
This work was supported in part by the Intel Science and Technology Center for Big Data and NSF CDI grant IIA-1028195.